\documentclass[10pt]{article}
\usepackage{natbib}
\usepackage{graphicx}
\usepackage{verbatim}
\usepackage{verbatim,float,url,enumerate}

\newtheorem{theorem}{Theorem}[section]
\newtheorem{lemma}[theorem]{Lemma}

\def\protolasso {{\tt protolasso}}
\def\Protolasso  {{\tt Protolasso}}

\def\prototest  {{\tt prototest}}
\def\Prototest {{\tt Prototest}}
\def \sI {{\cal I}}

\title{Sparse regression and marginal testing using cluster prototypes}
\author{Stephen Reid\thanks{Department of Statistics, Stanford
  University, sreid@stanford.edu   }   \hskip .1in and Robert Tibshirani\thanks{Departments of Health, Research \&
  Policy, and Statistics,
  Stanford University, tibs@stanford.edu}}
\date{}
\begin{document}
	\maketitle
	\begin{abstract}
	We propose a new approach for sparse regression and marginal testing, for data with correlated features.
	Our procedure first clusters the features, and then chooses  as the cluster prototype the most informative feature in that  cluster.
	Then we apply  either sparse regression (lasso)  or  marginal significance testing to  these prototypes.
	While this kind of strategy  is not entirely new, a key feature of our proposal is its use of the post-selection  inference theory
	of \citet{TTTPostSel} and \citet{LeeSun2TaylorPostSel}  to compute {\em exact} p-values and confidence intervals that properly account for the selection of prototypes.
	We also apply the recent ``knockoff'' idea  of \citet{BC2014} to provide exact finite sample control of the FDR of our regression procedure.
	We illustrate our proposals on both real and simulated data.
	\end{abstract}
	
		%%%% Introduction
	\section{Introduction}
	\label{sec:introduction}
	%%\marginpar{There may be places where Protolasso or protolass has no space after it, causing it to run into the next word. (I've tried to find all of these)  If you see this, add a backslash after protolasso. }
	We consider the linear model setup:
	\begin{equation}\label{linmod}
		y = X\beta + \epsilon
	\end{equation}
	where $y$ is the $n \times 1$ response vector, $X$ an $n \times p$ matrix of predictors, $\beta$ a $p \times 1$ vector of true regression coefficients and $\epsilon$ an $n \times 1$ vector of errors -- usually assumed to have $n$-dimensional Gaussian distribution $N(0, \sigma^2I)$. Although not critical for the ideas of this paper, we are especially interested in  the $p > n$ case. These types of predictor matrices are often encountered in practice and provide interesting case studies for our methods. Since our matrix is not of full column rank, least squares regression cannot be used for further analysis of the problem. Instead we proceed with a regression method that comprises of both variable selection and parameter estimation components. Examples include the lasso of \cite{TibsLasso}, the elastic net of \citet{enet}, principal component regression and forward and backward stepwise regression. We focus primarily on the lasso.
	
	Furthermore, suppose that the columns of $X$ share substantial empirical correlation -- artifacts of a hypothetical underlying covariance matrix for the columns that exhibits significant block structure. Such correlation amongst the columns of the predictor matrix poses a thorny set of problems, including numerical instability of regression solutions, parameter near-nonidentifiability and the inability of the lasso to recover the true set of signal variables. The lasso essentially selects randomly among a set of highly correlated signal variables, entering only one of them. Another, largely unaddressed, issue is that of a lack of interpretability. The lasso merely provides a set of predictive variables. It is silent on the grouping (correlation) structure amongst the variables or indeed the estimated effect on the response of those unselected variables. Of course, this tool is not designed for such a purpose.
	
	Our goal in this paper is to provide the owner of a wide dataset, exhibiting significant correlation amongst its columns, with a procedure to discover the column groupings and a single representative prototype from each group. Subsequent sparse regression or marginal testing is then  performed on these prototypes. Critically, our method (described in subsequent sections) allows us to leverage the powerful results of \citet{LeeSun2TaylorPostSel} and \cite{TTTPostSel},and subsequent work by \citet{LeeTaylorMC} to provide \textit{valid} confidence intervals and p-values (testing for nullity) of the effect sizes of these prototypes. This is true even after their selection as prototypes \textit{and} their participation in subsequent sparse regression. As an additional feature, we can use the group structure learned in the first step and generate similar confidence intervals and p-values for those variables not selected, but correlated with, the prototypes. Note that this provides a highly interpretable snapshot of the data: a grouping structure, effect size estimates of most predictive prototypes of the grouping structure as well as for the also-rans in each of these groups.
	
		Our proposal  comprises of a series of distinct steps:

	\begin{enumerate}
		\item {\bf Grouping}  the columns of the predictor matrix $X$ (assumed fixed), using either pre-defined  groups from the problem context  or a clustering method.
				\item {\bf Prototype extraction}, one from each cluster, by screening on the marginal correlation each cluster member has with the response $y$.
		\item{\bf  Subsequent regression analysis} on the selected prototypes, here specifically the lasso, what we call the \Protolasso,  or {\bf marginal testing} of
		the prototypes, what we call \Prototest. We use  the theory of {\em post-selection inference} to obtain { \em exact } p-values and confidence intervals that properly account for the selection at every stage.
	\end{enumerate}

	\begin{figure}[hbtp]
		\centering
		\includegraphics[width=120mm]{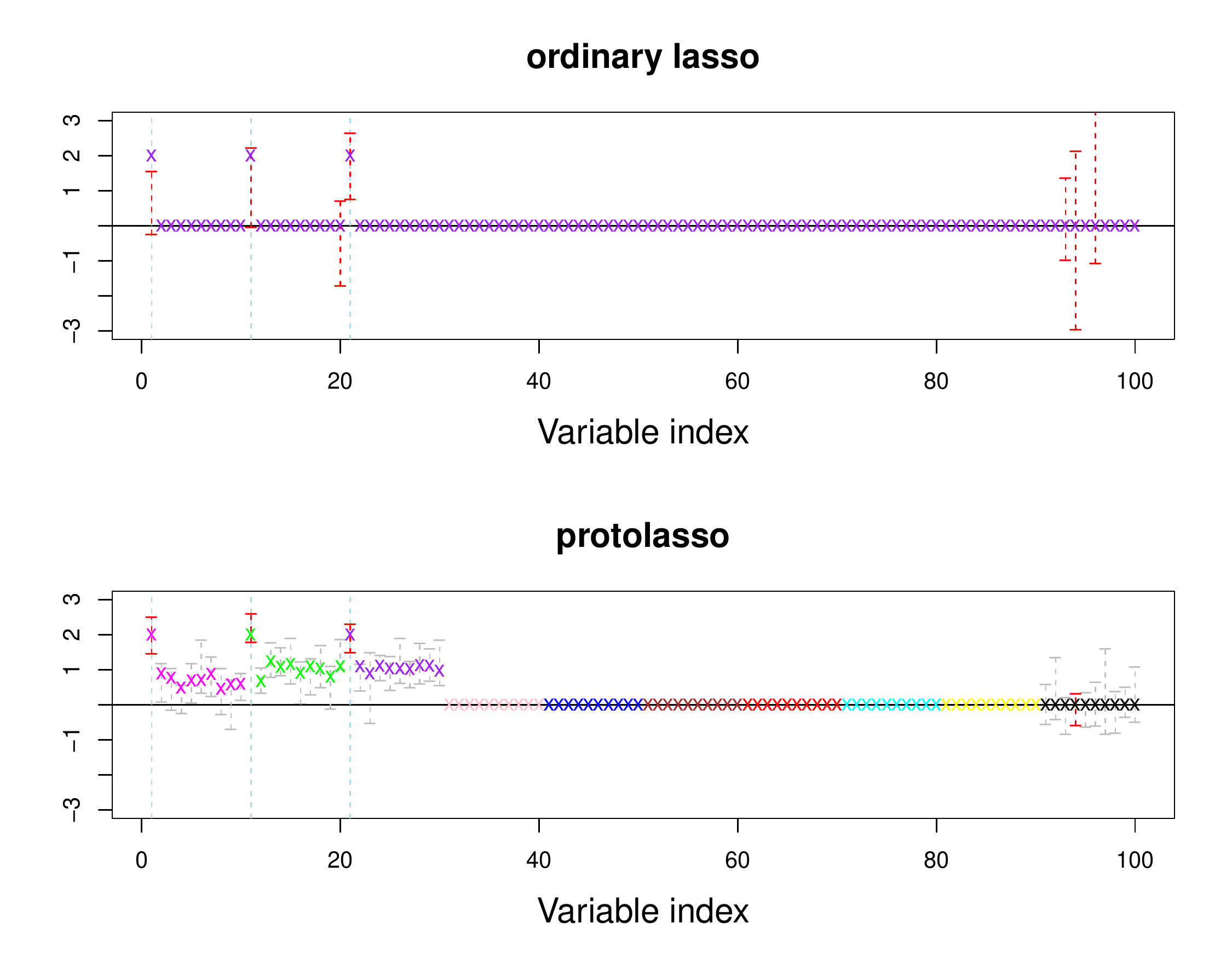}
		\caption{\emph{Confidence interval plots for selected variables from ordinary lasso (\textbf{top panel}) and {\tt protolasso} (\textbf{bottom panel}). Predictor matrix $X$ has $n = 50$ and $p = 100$ columns, generated to occur in groups of size 10 (columns 1 to 10, 11 to 20, 21 to 30, etc.). Predictors in each group share pairwise correlations of size $\rho = 0.5$ amongst themselves, but are uncorrelated with all other predictors. Crosses represent the true target of the confidence interval (the partial correlation of the appropriate predictor with the response, given the selected model). Cross colours represent the grouping detected by each method. The ordinary lasso does not detect groups, so all crosses have the same colour. Confidence intervals are represented by dashed vertical bars and horizontal endpoints. Red confidence bars represent those for the selected predictors/prototypes. grey confidence bands represent intervals derived by swapping out the group prototype for the variable of interest (described later). The ordinary lasso does not tell us anything about group structure, precluding the swapping of other group members for the prototype, hence all the confidence intervals are in red. If predictors/clusters prototypes were not selected, we do not construct a confidence interval for them -- hence the empty spaces. Dashed blue vertical lines show the indices of variables that were given non-zero signal in the original $\beta$ vector, with parameter value set to 2.}}
		\label{fig:egCI}
	\end{figure}
	
	As an illustration of the interpretive richness of our method, consider Figure~\ref{fig:egCI}. The caption describes how data was generated. A single pair of $X$ and $y$ was generated. To this data we fit the ordinary lasso, used cross-validation to select the optimal number of variables and used the post-selection inference tools of \citet{LeeSun2TaylorPostSel} to generate confidence intervals for the selected variables. We also subjected the data to our  \protolasso\  procedure. We describe this procedure in later sections. Some observations:
	
	\begin{itemize}
		\item Ordinary lasso selects seven variables and constructs confidence intervals for them. The three true signal variables (1, 11 and 21) are selected. 
		\item Confidence intervals for the ordinary lasso seem rather wide, especially when comparing to the lower panel. 
		\item The ordinary lasso does not detect any group structure in the predictors. All crosses in the top panel are of the same colour.
		\item \Protolasso\ detects the group structure in the columns (in fact, recovers it exactly). Notice the different colour crosses in the lower panel.
		\item \Protolasso\ selects four prototypes (variables 1, 11, 21 and 94) and constructs confidence intervals for them (red confidence interval bands), capturing all three true signal variables. Notice that the confidence intervals seem much narrower.
		\item The grouping structure captured by \protolasso\  allows us to construct confidence intervals for the non-prototypes in the groups of the selected prototypes. These are in grey. One can then ascertain the effect of replacing the prototype by the non-prototype of interest. Note that these confidence intervals are still valid and take the entire selection procedure into account. Clearly such an analysis is impossible when using ordinary lasso.
	\end{itemize}
	
	In subsequent sections, we discuss some related work, the detection of the group structure (first phase) and the tools used in the construction of our post selection confidence intervals and p-values. We conclude with some simulations and applications to real datasets.
	
	This paper is organized as follows. In section \ref{sec:related} we review some related work in the literature.
	Section \ref{sec:clustering} describes the clustering of the features and prototype extraction. In Section \ref{sec:post_sel_inf} we review the theory of post-selection inference and give details and examples of
	our \Protolasso\   proposal. The  \Prototest\   of  Section \ref{sec:prototest} carries out marginal testing of the selected prototypes.
	We discuss FDR control for \protolasso\    via knockoffs in Section \ref{sec:fdr}.  Appendix A gives details of  the proposed gap statistic for choosing the number of clusters.
	We conclude with discussion in Section \ref{sec:discussion}.
	\section{Related work}
	\label{sec:related}
	The lasso has become a widely used tool for linear regression with simultaneous variable selection. It can be applied to matrices with $p > n$. There has been much theoretical development of the method's variable selection and screening properties, under a variety of assumptions. References include  \citet{MeinhausenBuhlmann2006}, \citet{ZhaoYu2006}, \citet{vanDeGeer2007}, \citet{ZhangHuang2008}, \citet{vanDeGeerBuhlmann2009}, \citet{MeinhausenYu2009}, \citet{Bickel2009} and \citet{SunZhang2011}.
	
	These positive findings are heartening. However, it has been noted that the lasso performs poorly in the presence of predictors with high empirical correlation. In particular, it selects but a single variable into the model from a set of mutually highly correlated variables, even if the omitted ones are true signal variables. Other methods like the \textit{elastic net} of \citet{enet}, \textit{OSCAR} of \citet{oscar2008} and the \textit{clustered lasso} of \citet{She2010} have been developed to address this shortcoming. These offer improvements, but do not explicitly take the correlation amongst the variables into account. All of the methods consider different penalty terms in order to encourage more acceptable behavior in the face of highly correlated predictors.
	
	Another set of methods cluster the variables first and then fit a model. Some methods do these two steps sequentially (clustering first and then fitting a model to some cluster representatives); others simultaneously. Examples of the former include principal component regression, \textit{gene shaving} of \citet{geneshaving2000}, \textit{tree harvesting} of \citet{treeharvesting2001} and the canonical correlation clustering and subsequent sparse regression of \citet{buhlmannCorrelatedClustering}. The method of \citet{dettlingBuhlmann2004} and \textit{OSCAR} represent the latter.
	
	The above methods all attempt to find clusters of variables and combine variables within a cluster to support good predictive ability in the subsequent (or concomitant) model fit. Our method also proceeds sequentially by first clustering variables into highly correlated groups. A second step differentiates our method from the others: once the clusters have been found, we extract a single representative prototype from the cluster membership. We do not combine all members of the cluster by averaging or projecting onto principal component directions. The prototypes are chosen so as to preserve subsequent predictive ability as much as possible. 
	
	Once we have constructed clusters and extracted prototypes, the user can proceed as they wish. We, however, follow \citet{buhlmannCorrelatedClustering} and perform a sparse regression after prototype selection. The novelty of our method (and the source of its rich interpretability) is our liberal use of the post selection inference framework of  \cite{TTTPostSel}  and \citet{LeeSun2TaylorPostSel}. Our clustering, prototyping and subsequent sparse regression procedures are all chosen so as to ensure adherence to their framework, enabling the computation of a set of confidence intervals and p-values meant to give deeper understanding of the grouping structure, the significant prototypes and their potential doppelg\"{a}ngers. We discuss the clustering algorithm next.
	
	\section{Clustering and prototype extraction}
	\label{sec:clustering}
	The first step in our procedure is the estimation the grouping structure of the features.  In some problems these groupings may be pre-defined, for example, gene sets in microarray studies, organizing genes into functional units. In this case our procedure will make use of these groups.
	
	In many cases, however, no \textit{a priori} grouping is available for the features and 
	unsupervised clustering tools can be applied.  Fortunately, the post selection results of and \citet{LeeSun2TaylorPostSel} (the backbone of the ultimate interpretability of our method) all go through, conditional on the predictor matrix $X$. Any clustering performed on the columns of $X$ need not be taken into account when constructing these inferences later on, provided the clustering is truly unsupervised and receives no input from the response $y$. 
	
	As such, we do not prescribe to the user any particular clustering method. We are free to use $k$-means, $k$-medoids or any of the hierarchical clusterings: complete, single and minimax linkage (\citet{BienTibs2011}), for example. We proceed with hierarchical clustering methods.
	Minimax clustering has the added advantage of returning a clustering of the variables and a prototype for each cluster. These prototypes are determined solely on merit of $X$, with no input from $y$ and so may lack some predictive power. They could be used immediately in further analysis, but we propose a different set of prototypes, that can be extracted from the output of any clustering method. They are described in a subsequent paragraph. 

	Since we are concerned with empirical correlation, we prescribe that the dissimilarity matrix input to these clustering methods be $D$, a $p \times p$ matrix with $(j,k)^{th}$ entry:
	\[
		D_{jk} = 1 - \frac{\sum_{i = 1}^n(x_{ij} - \bar{x}_j)(x_{ik} - \bar{x}_k)}{\sqrt{\sum_{i = 1}^n(x_{ij} - \bar{x}_j)^2}\sqrt{\sum_{i = 1}^n(x_{ik} - \bar{x}_k)^2}}
	\]
	i.e. one minus the sample correlation of columns $j$ and $k$ where $\bar{x}_j = \frac{1}{n}\sum_{i = 1}^nx_{ij}$.
	
	Of course, each of the clustering algorithms require the user to set the number of required clusters as a parameter. Since our method forms somewhat of a pipeline of analyses, we do not want this step to be a bottleneck. We propose a method for the automatic selection of the number of clusters. Our proposal is  a gap statistic procedure very similar to that of \citet{gapstat}. To maintain the flow of the  exposition, we defer details of this procedure to Appendix A.	
		\subsection{Prototype extraction}
	\label{sec:prototype_extraction}
	Having identified clusters, the next step is to extract prototypes -- one from each cluster. By ``prototype" we mean a single cluster representative, chosen from \textit{among the members} of the cluster. We exclude the possibility of using a point wise mean or first principal component of the points in the cluster. 
	
	Some clustering methods produce such prototypes as a side effect of the algorithm. Examples of this are $k$-medoid clustering and hierarchical clustering with minimax linkage, as developed by \citet{BienTibs2011}. These methods produce prototypes in an unsupervised fashion, with no recourse to the response $y$. Although undoubtedly a good feature of these methods, our goal is to extract prototypes more explicitly tied to their correlation to the response. These in some sense represent the ``most interesting" members of the cluster when considering the response as well.
	
	Once we make use of $y$ in our prototype selection, and proceed with these prototypes in further analysis, any inference downstream must take this selection effect into account. Luckily, we can leverage the post selection inference framework of \citet{LeeSun2TaylorPostSel} to account for this selection effect, provided the prototype selection method is simple enough, i.e. linear in $y$, after conditioning on minimal additional information. One possibility would be to use the
``least squares prototype''--- that is, regress $y$ on the features in the cluster and used the fitted values as the ``prototype''. This would  capture the joint signal in all of the cluster members, and would be particularly appropriate when
modeling  genesets in genomic studies.
	
	However in  this paper we focus on simple marginal correlation screening. In particular, given a clustering with $K$ clusters, written as $\{C_1, C_2, \dots, C_K\}$ where $\cup_{k = 1}^KC_k = \{1, 2, \dots, p\}$ and $C_j \cap C_k = \phi$ for $j \neq k$, we extract prototypes $\hat{P} = \{\hat{P}_1, \hat{P}_2, \dots, \hat{P}_K\}$ where each $\hat{P}_k \in \{1, 2, \dots, p\}$ and $\hat{P}_k \in C_k$ such that
	\[
		\hat{P}_k = {\rm argmax}_{j \in C_k}|x^\top_jy|
	\]
	
	It turns out that the selection of prototypes in this way is easily translated into the framework of \citet{LeeSun2TaylorPostSel} and hence easily incorporated into subsequent inference. Even at this juncture, it is possible to make inferences (perform significance tests and construct confidence intervals) for these prototypes. Particular examples include inferences on the individual marginal correlations with $y$ and partial correlations if all (or some) of the prototypes are used in a subsequent least squares regression. Details of post selection inference are discussed in the next section.
	
	\begin{figure}[hbtp]
		\centering
		\includegraphics[width=130mm]{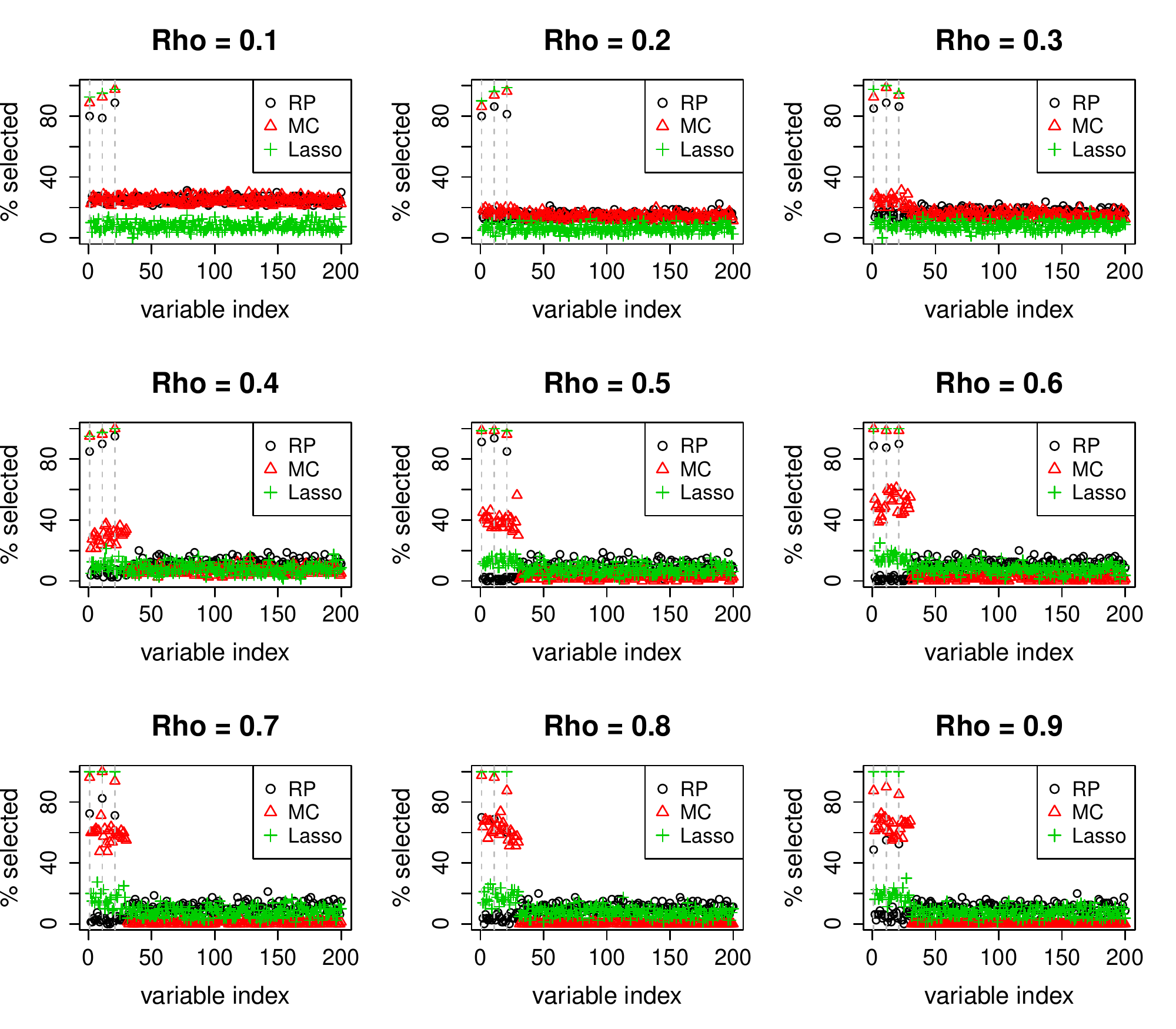}
		\caption{\emph{Dataset has $n = 50$ rows and $p = 200$ columns, divided into 20 groups of 10 columns each. Columns within a group share pairwise correlation $\rho$, which is set to $\rho = 0.1, 0.2, \dots, 0.9$ from the top left to the bottom right. Horizontal axes measure the variable number, while vertical axes measure the proportion of times out of $S = 80$ simulation runs in which each variable was selected by each of the methods. Black circles represent the \protolasso\ procedure, red triangles the marginal screening procedure and the green plusses the ordinary lasso fit to all the variables. $\beta$ has 3 non-zero entries -- all set to 2 -- at variables 1, 11 and 21. These are indicated by the three vertical grey dashed lines.}}
		\label{fig:var_sel}
	\end{figure}
	
	Figure~\ref{fig:var_sel} demonstrates that the variable selection performance of the \protolasso \newline
	procedure (clustering and then marginal correlation screening in each cluster) is not much worse than that of two other selection procedures. The first of the other methods is a marginal screening method, each time selecting as many variables as there are clusters in the \protolasso\ method, retaining those with the largest absolute marginal correlation with $y$. The second is the lasso, applied to the entire dataset, with the regularization parameter chosen so as to ensure the same number of variables is selected.
	
	The lasso performs admirably, over all correlations -- a function of the sheer size of the signal. The \protolasso\ procedure is not far behind, selecting the non-prototypes in the signal clusters far less often than does the other two methods, while selecting some of the noise variables more often. This is a function of selecting a prototype from each cluster, many of which are filled only with noise variables. Selection performance degrades at very high correlations. However, at these correlations it is very difficult to distinguish between signal and non-signal variables in the same cluster anyway.
	
	The reader should note that our method is not designed to be competitive in the selection of signal variables. There are other methods more effective at this endeavor. The merit of our method is its rich interpretability. We show the output of this particular simulation merely to demonstrate that variable selection performance is not too critically compromised.
	
	\section{Post selection inference}
	\label{sec:post_sel_inf}
	
\citet{TTTPostSel}  and 	\citet{LeeSun2TaylorPostSel} present a powerful framework for inference after variable selection that takes the response into account. They consider the usual linear model, as in Equation~(\ref{linmod}), assuming in particular that $\epsilon \sim N(0, \Sigma)$. Let $\mu = X\beta$. 
	
	Their focus is inference on linear combinations $\eta^\top\mu$ \textit{after} the application of a variable selection technique. Variable selection -- a process usually privy to the response $y$ -- complicates post fit inference. Given that a set of variables $\hat{M} \subset \{1, 2, \dots, p\}$ has been selected (often with $|\hat{M}| \leq n$), they argue correctly that the usual Gaussian least squares inference on $\beta_{\hat{M}} = X_{\hat{M}}^+\mu$, is invalid. Here the subscript denotes the restriction of the columns of $X$ (or the entries of $\beta$) to those in the set $\hat{M}$. The superscript $+$ denotes the Moore-Penrose of the matrix in question.
	
	Their main results pertain to the lasso. They show that, for fixed $X$, and conditioning on both the variables selected by the lasso $\hat{M} = M$ and the signs $s_{M}$ of the non-zero $\beta$,  we can construct matrices $A = A^{lasso}_{M, s}$ and $b = b^{lasso}_{M, s}$ such that the selection event $\{\hat{M} = M, s_{\hat{M}} = s_{M}\}$ can be written as a set of \textit{linear} constraints on $y$:
	\[
		A^{lasso}_{M, s_M}y \leq b^{lasso}_{M, s_M}
	\]
	Here $s_{M, i} = {\rm sign}(\beta_{M, j})$. The authors give explicit formulae for $A$ and $b$ in their paper. The reader is referred there for details. 
	
	This result can then be used to do inference on $\eta^\top\mu$ post selection, by considering the distribution of:
	\[
		\eta^\top y | Ay \leq b
	\]
	 
	This allows  us to characterize the conditional distribution exactly:
	\[
		F^{[\mathcal{V}^-, \mathcal{V}^+]}_{\eta^\top\mu, \eta^\top\Sigma\eta}(\eta^\top y) | Ay\leq b \sim {\rm Unif(0, 1)}
	\]
	where
	\[
		F^{[a,b]}_{\mu, \sigma^2}(x) = \frac{\Phi\left(\frac{x-\mu}{\sigma}\right) - \Phi\left(\frac{a-\mu}{\sigma}\right)}{\Phi\left(\frac{b-\mu}{\sigma}\right) - \Phi\left(\frac{a-\mu}{\sigma}\right)}
	\]
	the truncated Gaussian cumulative distribution function (cdf). Again, the exact forms of $\mathcal{V}^-$ and $\mathcal{V}^+$ are given in the paper.
	
	Note that the true utility of this result extends beyond a mere application to the lasso selection procedure. Indeed, any selection procedure that, with perhaps a little extra conditioning information, can be written as a set of linear constraints on $y$ can be subjected to this framework. Other examples of such procedures are marginal correlation screening, as described in \citet{LeeTaylorMC} and, in the multivariate mean vector estimation literature, the Benjamini-Hochberg procedure and the selection of the largest $K$ elements of $y$, as described in \citet{reidManyMeans}.
	
	For a given procedure, the challenge is to write the selection constraints it imposes as a set of linear constraints on $y$ (if possible). This gives one the forms of $A$ and $b$ specific to that method. Coupled with a particular contrast of interest $\eta$, which can be chosen dependent on the selected variable set $\hat{M}$, one can then find particular values for $\mathcal{V}^-$ and $\mathcal{V}^+$. With the addition of observed response data $y$ and an assumption of known variance $\Sigma$, the distribution of the above is fully characterized, allowing p-values and confidence intervals to be computed via evaluations or inversions of the truncated Gaussian cdf. 
	
	Our method of clustering, prototyping and subsequent sparse regression on prototypes is easily cast into this framework. In fact, the same framework can be used to do inference even on non-prototypes, leading to richer interpretation of the dataset. This is discussed next.
	
	\subsection{Details of  post selection inference for \Protolasso}
	\label{subsec:protolasso}
	
	Recall the  steps of  the \protolasso\ and \prototest\ procedures  given in the Introduction, and summarized here for reference:
	\vskip .25in
	\centerline {{\bf The \Protolasso\ and \Prototest\ procedures}}
	
	\begin{enumerate}
	\item {\em Group the features}: use either a pre-defined grouping of features, or  cluster the features into groups.
	\item {\em Form prototypes}: extract a prototype feature from each cluster, by choosing the feature  with highest marginal correlation with the response $y$;
	\item {\em Model fitting and inference:}  carry out a regression analysis  on the selected prototypes (\protolasso) or {marginal testing }of the prototypes (\prototest).
	Use the theory of post-selection inference to compute exact p-values and confidence intervals that account for the selection effects.
		\end{enumerate}
		\vskip .25in
	We assume the predictor matrix $X$ fixed. This is also the assumption in the post selection inference framework discussed above. Note that if one assumes $X$ fixed and if the clustering performed on the columns is completely unsupervised (including the method used to estimate the number of clusters), then the post selection distributional results remain unaffected. Identifying the grouping structure in the fixed $X$ has no effect on the linear constraints on $y$.  A new replication of $y$ will \textit{not} change the grouping structure so identified. The first step need not be accounted for in the post selection framework.
	
	The second step does involve $y$ and thus contributes some linear inequalities to the post selection inference. Suppose we are given a cluster $C_1$, its prototype $\hat{P}_1$ and the sign $\hat{s}^{(1)}_1 = {\rm sign}(x_{\hat{P}_1}^\top y)$. According to \citet{LeeTaylorMC}, the constraints contributed by selecting the prototype in this cluster can be represented by the matrices
	\begin{equation}
		A^{(1)}_{\hat{P}_1} = \left( \begin{array}{c} X_{C_1 \setminus \{\hat{P}_1\}}^\top -  s^{(1)}_1\textbf{1}X_{\hat{P}_1}^\top \\ -X_{C_1 \setminus \{\hat{P}_1\}}^\top -  s^{(1)}_1\textbf{1}X_{\hat{P}_1}^\top\end{array}\right)
		\label{eqn:A1}
	\end{equation}
	\begin{equation}
		b^{(1)}_{\hat{P}_1} = \left(\begin{array}{c} \textbf{0} \\ \textbf{0} \end{array}\right)
		\label{eqn:b1}
	\end{equation}
	Collecting the constraints for all the clusters, the entire prototyping step contributes constraint matrices
	\begin{equation}
		A_{\hat{P}}^{(1)} = \left(\begin{array}{c} A^{(1)}_{\hat{P}_1} \\ A^{(2)}_{\hat{P}_2} \\ \vdots \\ A^{(K)}_{\hat{P}_K}  \end{array}\right)
		\label{eqn:A2}
		\end{equation}
	and
\begin{equation}
		b_{\hat{P}}^{(1)} = \left(\begin{array}{c} b^{(1)}_{\hat{P}_1} \\ b^{(2)}_{\hat{P}_2} \\ \vdots \\ b^{(K)}_{\hat{P}_K} \end{array}\right)
		\label{eqn:b2}
		\end{equation}
	where we condition on the set of selected prototypes and the signs of their marginal correlations with $y$, i.e. the event $\{\hat{P} = P, s^{(1)} = s\}$ where $s^{(1)} = (s^{(1)}_1, s^{(2)}_2, \dots, s^{(K)}_K)$.
	
	Finally, the third step proceeds by fitting the lasso on the columns of $X_{\hat{P}}$. Following \citet{LeeSun2TaylorPostSel}, we can construct the constraint matrices for this regression for a fixed regularization parameter $\lambda$, once we condition on the set of indices with non-zero coefficients $\hat{M} \subset \hat{P}$ and the signs of these non-zero coefficients $\hat{s}^{(2)}$. On the conditioning event $\{\hat{P} = P, \hat{M} = M, \hat{s}^{(2)} = s^{(2)}\}$, we have the constraint matrices
	\[
		A^{(2)} = \left(\begin{array}{c} \frac{1}{\lambda}X^\top_{P \setminus M}(I - Q_{M}) \\ -\frac{1}{\lambda}X^\top_{P \setminus M}(I - Q_{M}) \\ -{\rm diag}(s^{(2)})\left(X^\top_{M}X_{M}\right)^{-1}X_M^\top\end{array}\right)
	\]
	\[
		b^{(2)} = \left(\begin{array}{c} \textbf{1} - X^\top_{P\setminus M}(X^\top)^+s^{(2)} \\ \textbf{1} + X^\top_{P\setminus M}(X^\top)^+s^{(2)} \\ -\lambda\cdot{\rm diag}(s^{(2)})(X^\top_MX_M)^{-1}s^{(2)} \end{array}\right)
	\]
	where $Q_M = X_M\left(X_M^\top X_M\right)^{-1}X_M^\top$.
	
	Overall then, if we condition on the event $\{\hat{P} = P, \hat{M} = M, \hat{s}^{(1)} = s^{(1)}, \hat{s}^{(2)} = s^{(2)}\}$, we have the post selection distribution
	\begin{equation} \label{eq:post_sel_dist}
		\eta^\top y \sim F^{[\mathcal{V}^-, \mathcal{V}^+]}_{\eta^\top\mu, \sigma^2 \eta^\top\eta}
	\end{equation}
	where $\mathcal{V}^-$ and $\mathcal{V}^+$ are gleaned from $A = \left(A^{(1)\top},A^{(2)\top} \right)^\top$, $b = \left(b^{(1)\top},b^{(2)\top} \right)^\top$, $s^{(1)}$, $s^{(2)}$ and $\eta$ as described in \citet{LeeSun2TaylorPostSel}. Assuming that $\sigma^2$ is known, we are free to do inference on $\eta^\top\mu$.
	
	\subsection{Inference for the selected prototypes}
	
	Suppose that we have run the \protolasso\ procedure and are interested in the inference of the partial correlation of selected prototype $\hat{P}_k$ with the response $y$ i.e. the regression coefficient for that prototype obtained from the regression of $y$ onto the selected prototypes. In this case, we choose $\eta = (X_{M}^\top)^+_j$, where $j$ is the index in $M$ associated with $\hat{P}_k$, giving us values for $\mathcal{V}^-$ and $\mathcal{V}^+$.
	
	P-values for this coefficient $\eta^\top\mu$ (when testing for nullity) are gleaned by evaluating the cumulative distribution function described in Equation~(\ref{eq:post_sel_dist}). Confidence intervals are obtained by inverting the same function, solving for $\eta^\top\mu$, given data $\eta^\top y$. Indeed, this is how we obtained the red confidence intervals in lower panel of Figure~\ref{fig:egCI}.
	
	\subsection{Inference for other members of selected clusters}
	We need not only limit ourselves to inference on the selected prototypes alone. The post selection inference framework allows us to say something about the other members of clusters of the selected prototypes. Suppose that the second stage lasso selected prototype $\hat{P}_k$ (amongst others). The first stage of the \protolasso\ procedure gives us an estimated group structure. It could be of interest to say something about the effect size of one of the other members of cluster $C_k$ when we swap it for the selected prototype, but keep the prototypes from other selected clusters in the model.
	
	Formally, suppose we have selected prototypes with indices in $\hat{M}$ and that $\hat{P}_k \in \hat{M}$. Furthermore, suppose that $C_k$ has elements $j_1, j_2, \dots, j_{|C_k|}$ and we want to exchange $j_1 \neq \hat{P}_k$ for the the prototype. We can perform inference in this case by merely selecting the correct form of $\eta$. In particular, let
	\[
		\eta = (X^\top_{\hat{M}\setminus \{\hat{P}_j\} \cup \{j_1\}})^+_j
	\]
	where again $j$ is the index in $\hat{M}$ corresponding to $\hat{P}_k$. Having defined an $\eta$, we use the $A$ and $b$ matrices determined earlier to find values for $\mathcal{V}^-$ and $\mathcal{V}^+$ and are then free to proceed with inference. This is how we obtained the gray (non-prototype) confidence intervals in the lower panel of Figure~\ref{fig:egCI}.
	
	\begin{figure}
		\centering
		\includegraphics[width=100mm]{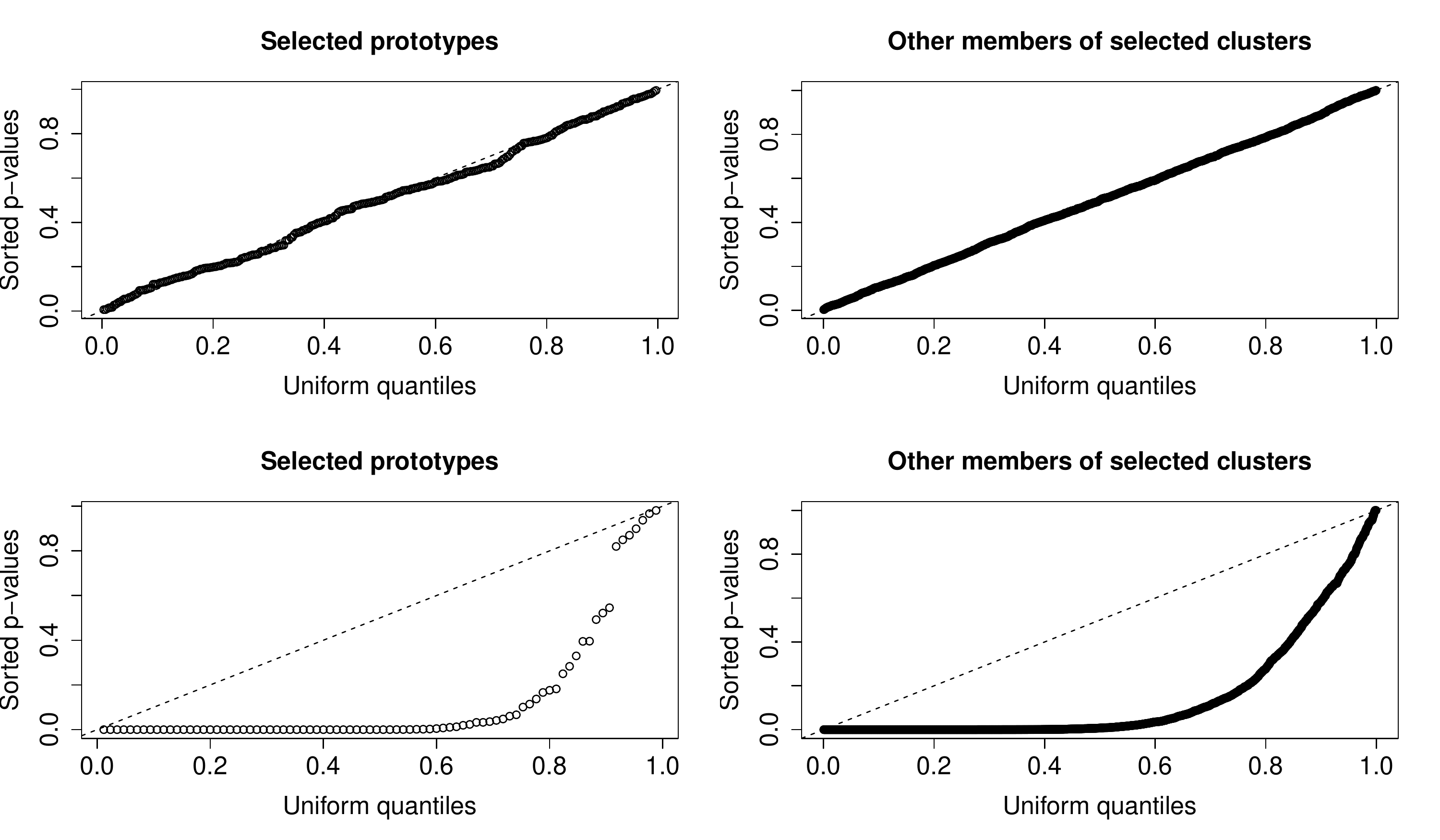}
		\caption{\emph{Dataset as in Figure~\ref{fig:egCI}. \textbf{Top row}: Sorted p-values (testing for nullity) versus uniform order statistics for selected prototypes (\textbf{left}) and other members of selected clusters (\textbf{right}) when $\beta = 0$. Notice the close concordance with the 45 degree line through the origin, as suggested by the theory. \textbf{Bottom row:} Same, but for $\beta_1 = \beta_{11} = \beta_{21} = 2$. Notice the sub-uniform behavior of the p-values.}}
		\label{fig:pvals}
	\end{figure}
	
	To convince the reader that these procedures are indeed valid, consider the dataset described in Figure~\ref{fig:egCI}. 
	
	%%\marginpar{Why ref  fig18 (so far ahead)? LOOKs STRANGE}
	
	We generated $S = 100$ replications of this dataset, each time constructing two responses: one with zero signal $\beta = 0$ and one with non-zero signal $\beta_1 = \beta_{11} = \beta_{21} = 2$. We used the gap statistic to select clusters and set the regularization parameter in the second step lasso to ensure that three prototypes were selected. We then computed p-values, under both signal regimes, for the selected prototypes (left panel Figure~\ref{fig:pvals}) and all other members of clusters with selected prototypes (right panel Figure~\ref{fig:pvals}). Notice that the p-values behave as one would expect -- they have uniform distribution for the zero signal ($\beta = 0$) case and a sub-uniform distribution when there is non-zero signal.

{\bf Remark.} We note that other supervised learning procedures like forward stepwise regression or LAR could be applied to the prototypes:  the post-selection inference theory is similar, and would  follow the  development  in \citet{TTTPostSel}	

	\section{Simulation study}
	\label{sec:simulation}
	In this section, we subject our method to a simulation study. We observe how its performance compares to that of the lasso and marginal screening (both without an initial clustering step). Performance is measured on datasets exhibiting some grouping structure. The correlation within groups is a variable we control to ascertain whether performance differs with changes in strength of the grouping structure.
	
	\subsection{The setup}
	At each setting of the simulation parameters, we generate a single matrix $X$ once. It has $n = 50$ rows and $p = 100$ columns. We consider two different kinds of grouping structures (discussed presently). Intra group correlation is set to $\rho = 0.1, 0.4$ and $0.7$ respectively for all pairs of variables within a group. Different intra group correlations allow us to simulate varying degrees of grouping structure strength. Inter group correlation amongst variables is zero.
	
	Once we have generated the predictor matrix (centered and scaled so that each column $x_j$ is such that $||x_j||_2 = n$), we generate $S = 100$ replications of the response $y$ according to Equation~(\ref{linmod}), with $\sigma = 1$.
	
	We consider two types of grouping structures, each with their own configurations for $\beta$:
	\begin{enumerate}
		\item \textit{Block diagonal}: The $p = 100$ variables are divided into 10 groups, each of size 10. Three $\beta$ configurations are considered:
		\begin{enumerate}
			\item \textit{Single signal per cluster}: $\beta_1 = \beta_{11} = \beta_{21} = \beta_{31} = \beta_{41} = \beta_{51} = \beta_* \neq 0$, with all other $\beta$ entries zero. $\beta_*$ is set at our discretion with its current value always explicitly stated in the relevant output.
			\item \textit{Paired signals}: $\beta_1 = \beta_2 = \beta_{11} = \beta_{12} = \beta_{21} = \beta_{22} = \beta_*$. Note that this configuration might confuse the lasso for larger configurations.
			\item \textit{Tight signal}: $\beta_1 = \beta_2 = \beta_3 = \beta_4 = \beta_5 = \beta_6 = \beta_*$. This could be even more confusing for the lasso for larger correlations.
		\end{enumerate}
		\item \textit{Single block}: There is one large block of correlated variables numbered from 1 to 50. All other variables are uncorrelated (i.e. blocks of size 1). Two $\beta$ configurations are considered:
		\begin{enumerate}
			\item \textit{Tight signal}: $\beta_1 = \beta_2 = \beta_3 = \beta_4 = \beta_5 = \beta_6 = \beta_*$.
			\item \textit{Split signal}: $\beta_1 = \beta_2 = \beta_3 = \beta_{51} = \beta_{52} = \beta_{53} = \beta_*$.
		\end{enumerate}
	\end{enumerate}
	In the interest of saving space, we do not present all output for all configurations. Often performance follows a similar theme for different configurations and only one needs to be presented in detail.
	
	\subsection{Performance measures}
	The utility of our method hinges on interpretability and post-selection inference. As such, we consider two broad metrics of success: \textit{variables entertained} and \textit{confidence interval width}. 
	
	We have already alluded to how our method can be used to make inference about not only the selected prototypes, but also those variables within the same cluster, once they are switched in for the prototype. Even if not directly selected, such a variable can still be entertained for further analysis. 
	
	Each $\beta$ configuration defines a set $S_0$ of indices corresponding to non-zero entries. If our method is to be successful, we wish is to return an ``entertained set" $\hat{S}$ that captures as much of $S_0$ as possible. We define $\hat{S}$ as the set of prototypes selected in the second stage lasso fit \textit{and} all the members of the clusters of these prototypes. The entertained sets for the other two methods are the set of indices corresponding to non-zero coefficient estimates (for the lasso) and the largest marginal correlations with $y$ (for marginal screening) respectively. The absence of an estimated group structure precludes the addition of any further elements to the entertained sets of these latter two methods.
	
	A method generates an entertained set $\hat{S}$ and its \textit{entertained proportion} is defined to be
	\[
		EP = \frac{|S_0 \cap \hat{S}|}{|S_0|}
	\]
	For each simulation run, $b$, we obtain the entertained proportion for each of the three methods: $EP^{pl}_b$, $EP^{lasso}_b$ and $EP^{ms}_b$. Since interest focuses on the relative performance of our method, we consider the measures:
	\[
		\frac{1}{S}\sum_{b = 1}^S (EP^{pl}_b - EP^{lasso}_b)
	\]
	\[
		\frac{1}{S}\sum_{b = 1}^S (EP^{pl}_b - EP^{ms}_b)
	\]
	These are plotted in Figures~\ref{fig:ep_heat1},~\ref{fig:ep_heat2} and~\ref{fig:ep_heat3}, for example. Notice that these measures vary between -1 and 1, with larger (more positive values) implying that the \protolasso\ procedure entertains a greater proportion of true variables on average.
	
	We also measure the confidence interval width of the confidence intervals constructed for each of the methods. Each method produces a set $\hat{M}$ of selected variables (here only the selected prototypes for the \protolasso\ procedure). The post selection inference framework allows us to construct intervals for the elements of $(X^\top_{\hat{M}})^+\mu$. Each method produces its own $A$ and $b$ matrices. For each replication, we compute the confidence intervals for all the elements of $\hat{M}^{pl}$ (only selected prototypes), $\hat{M}^{lasso}$ and $\hat{M}^{ms}$, take the median interval width over these indices, giving $MW_b^{pl}$, $MW_b^{lasso}$ and $MW_b^{ms}$, where ``MW" is short for ``median width". Finally, we consider the measures
	\[
		\frac{{\rm median}_b MW_b^{lasso}}{{\rm median}_b MW_b^{lasso} + {\rm median}_b MW_b^{pl}}
	\]
	\[
		\frac{{\rm median}_b MW_b^{ms}}{{\rm median}_b MW_b^{ms} + {\rm median}_b MW_b^{pl}}
	\]		
	Note these measures vary between 0 and 1, with larger values implying that the \protolasso\ method tends to produce narrower intervals (which is obviously better than the alternative). These are plotted in Figures~\ref{fig:ci_heat1} and~\ref{fig:ci_heat2}, for example. We choose this specific form to ensure that the measure ranges within a limited, well defined range. Results are presented in a heat map representation. Colors are more easily mapped to limited range measures.
	
	Finally, we control the number of clusters extracted from the hierarchical clustering and adjust the regularization parameters in the subsequent sparse regression so as to fix the number of these clusters that are selected at the second stage. We control and vary these numbers, despite \protolasso\ having an automatic mechanism for deciding on these numbers. The control and variation of these parameters allow us to study the effect on our performance measures of allowing different numbers of variables into the analysis. At each replication, the preset number of clusters and selected clusters determine the entertained set of the \protolasso\ procedure. We then ensure that the other two methods (lasso and marginal screening) have the \textit{same sized entertained sets}, by adjusting the relevant regularization parameters. Methods can then be measured on equal footing. We considered numbers of clusters ranging between 8 and 12 and selected clusters ranging between 1 and 7.
	
	\subsection{Entertained sets}
	
	\begin{figure}[hbtp]
		\centering
		\includegraphics[width=90mm]{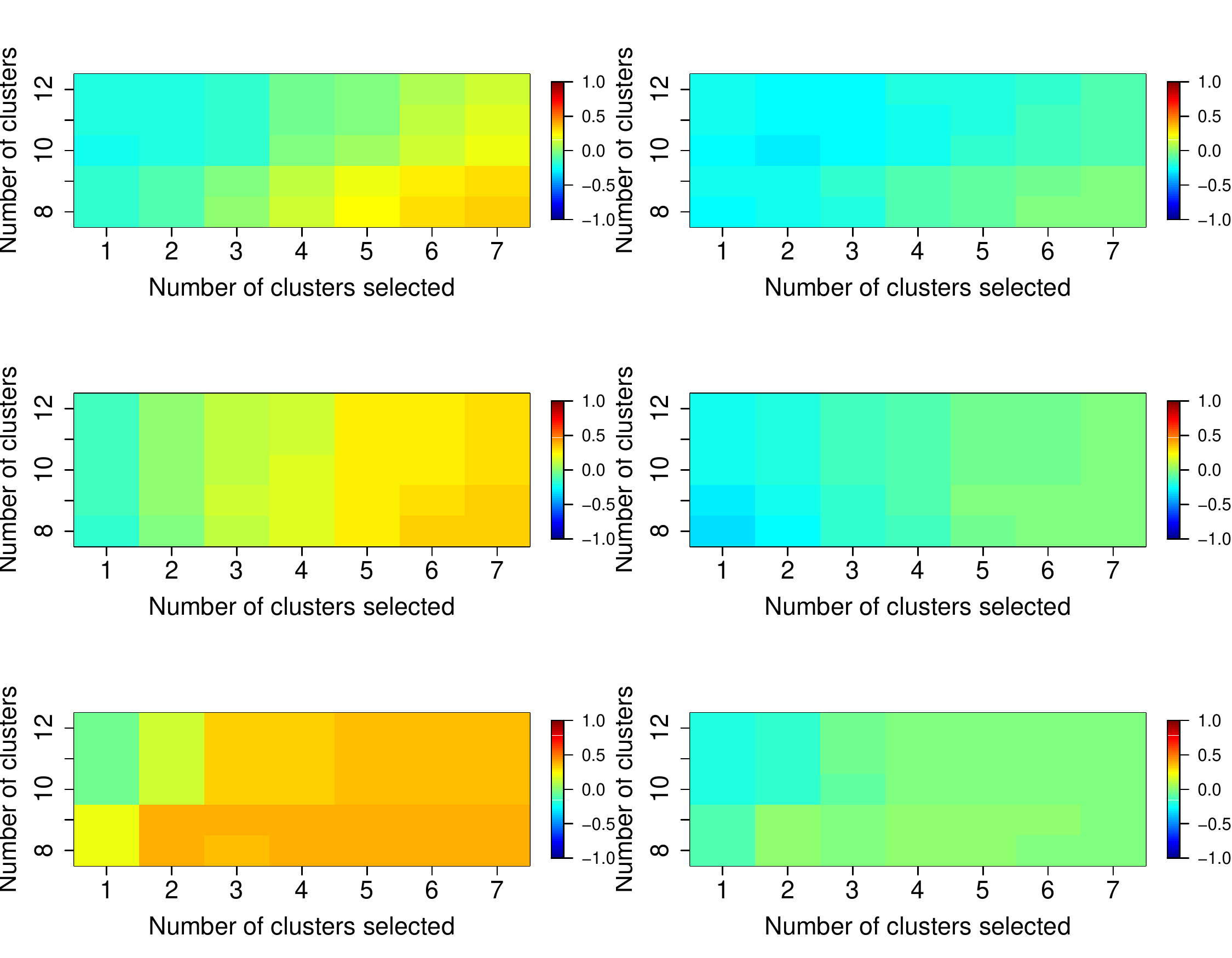}
		\caption{\emph{Heatmaps of entertained proportion performance measures as function of the initially extracted clusters (vertical axis) and those selected by the second stage lasso (horizontal axis). Dataset is the block diagonal $X$ with the paired signal $\beta$ configuration. $\beta_* = 0.2$. \textbf{Left panels: } Relative to lasso on all variables. \textbf{Right panels: } Relative to marginal screening. \textbf{Top row:} Pairwise correlation of variables in groups $\rho = 0.1$; \textbf{middle row:} $\rho = 0.4$; \textbf{bottom row:} $\rho = 0.7$. Colour scale on the right of each plot. ``More red" signifies that measure is closer to 1 (and \protolasso\ outperforms the relevant method), while ``more blue" signifies poorer performance relative to the other method.}}
		\label{fig:ep_heat1}
	\end{figure}
	
\begin{figure}[hbtp]
		\centering
		\includegraphics[width=90mm]{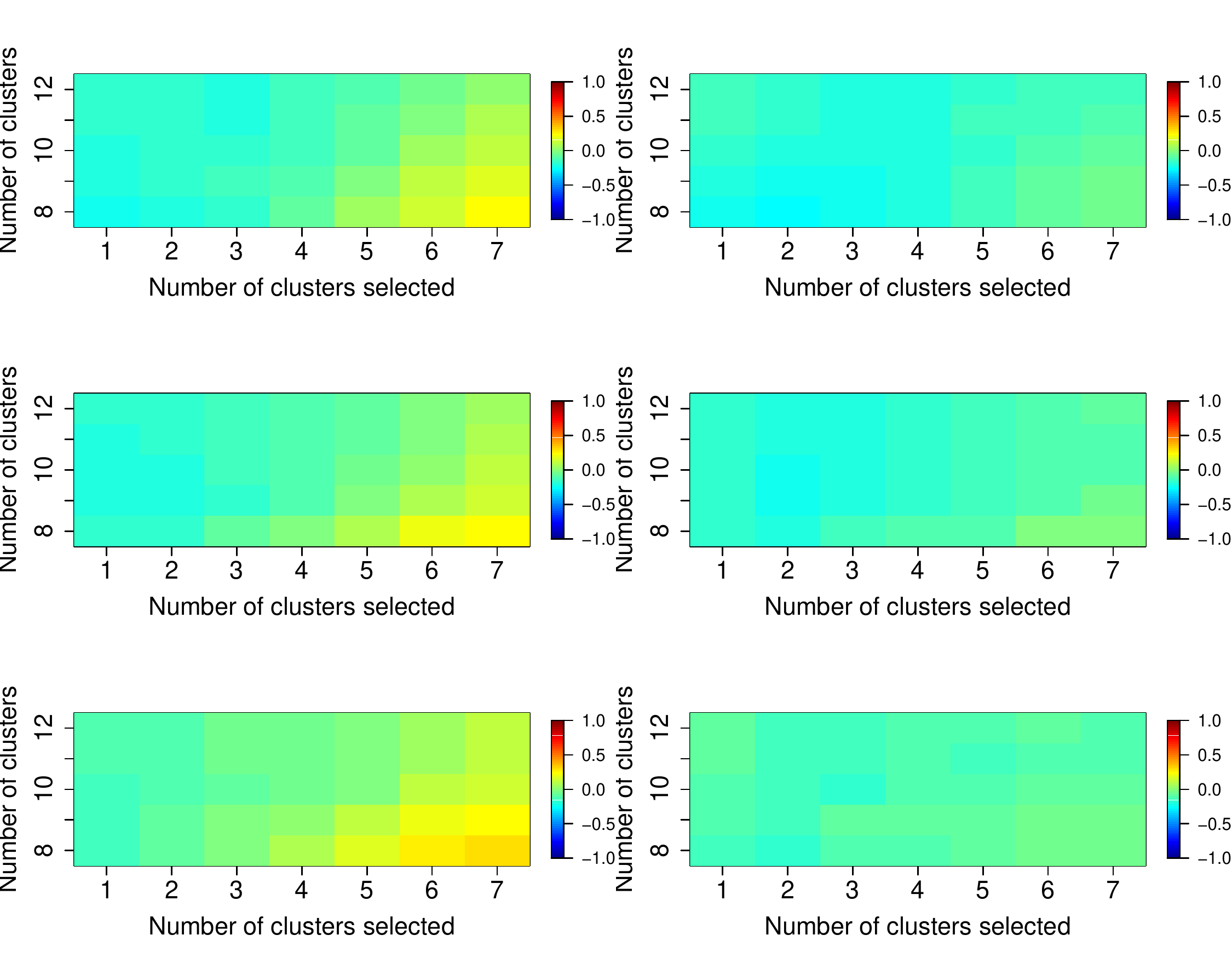}
		\caption{\emph{Heatmaps of entertained proportion performance measures as function of the initially extracted clusters (vertical axis) and those selected by the second stage lasso (horizontal axis). Dataset is the single block $X$ with the single signal $\beta$ configuration. $\beta_* = 0.2$. Same setup as Figure~\ref{fig:ep_heat1}.}}
		\label{fig:ep_heat2}
	\end{figure}

\begin{figure}[hbtp]
		\centering
		\includegraphics[width=90mm]{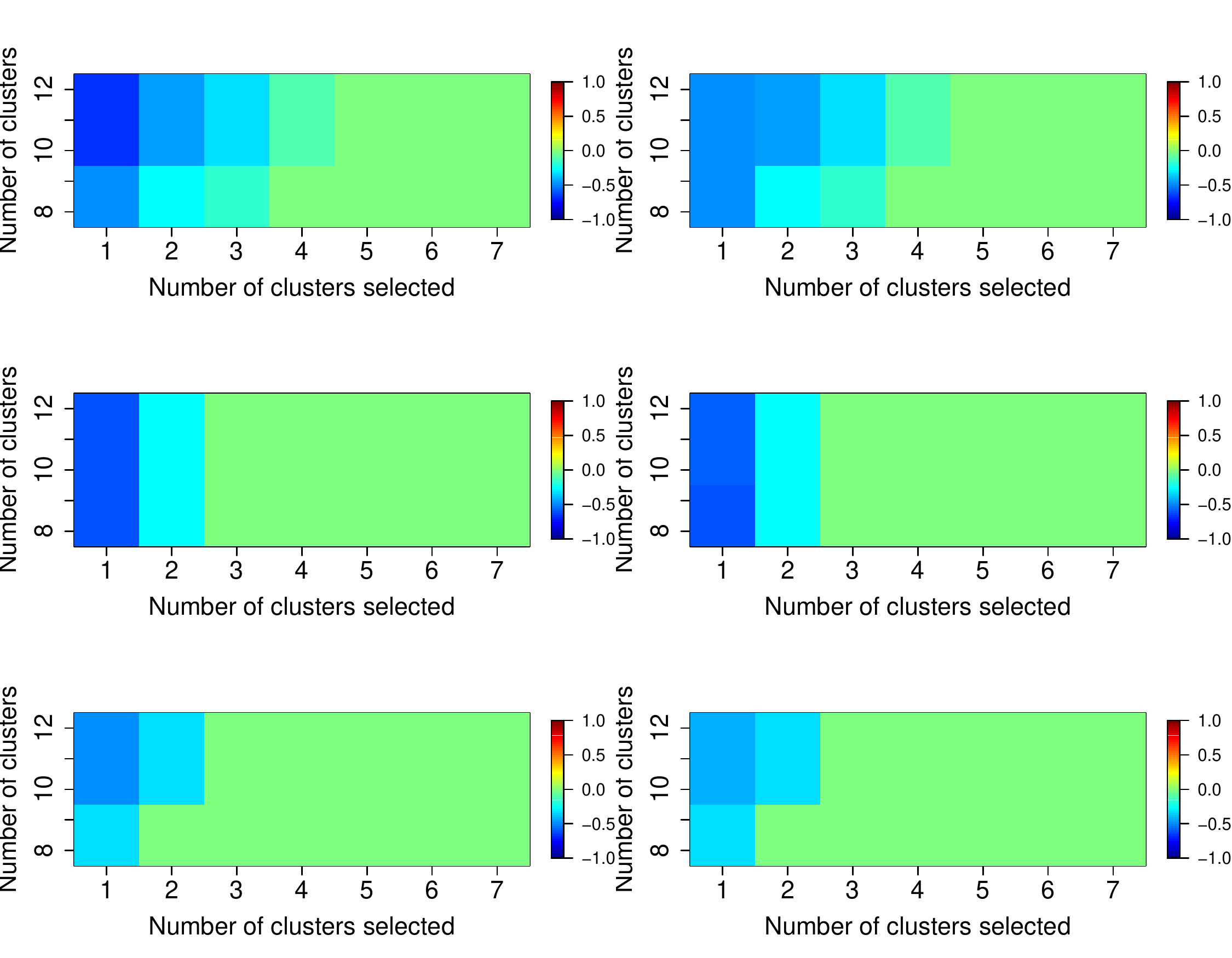}
		\caption{\emph{Heatmaps of entertained proportion performance measures as function of the initially extracted clusters (vertical axis) and those selected by the second stage lasso (horizontal axis). Dataset is the block diagonal $X$ with the paired signal $\beta$ configuration. $\beta_* = 2$. Same setup as Figure~\ref{fig:ep_heat1}.}}
		\label{fig:ep_heat3}
	\end{figure}

	Figures~\ref{fig:ep_heat1} and~\ref{fig:ep_heat2} show the entertained proportion performance of \protolasso\ against the other two methods for a reasonably small signal size $\beta_* = 0.2$. Some general observations:
	\begin{itemize}
		\item Entertained proportion performance is fairly similar for the \protolasso\ and marginal screening methods. We see this from the general green tinge in the right column. This regularity seems to persist over different correlations, dataset types and signal configurations. Perhaps some improvement of \protolasso\ as $\rho$ increases.
		\item Looking down the left panels, we see that \protolasso\ performance relative to the original lasso improves as $\rho$ increases. Recall that Figure~\ref{fig:ep_heat1} is for the paired signal configuration. There are pairs of ever more correlated variables, both with signal. The lasso tends to select one or the other, but rarely both and one or the other may not make it into the entertained set. \Protolasso, on the other hand, groups correlated variables together initially and tries to ensure that they do not both enter the second stage lasso. However, once we have selected any member of a cluster containing these signal variables, both enter the entertained set. \Protolasso\ then seems to counter the shortcoming of lasso by the nature in which it constructs its entertained set.
		\item This effect is attenuated for the single signal case in Figure~\ref{fig:ep_heat2}. Here we do not have correlated variables both getting signal. Each correlated cluster has at most one signal variable and the lasso is not so easily confused. Notice that \protolasso\ and lasso perform similarly. Although not shown, the tight signal cluster accentuates the effect of the previous bullet: \protolasso\ outperforms lasso to a greater extent, as all signal variables are correlated and in the same cluster.
		\item As we increase the signal size to $\beta_* = 2$ in Figure~\ref{fig:ep_heat3}, we see that all methods perform very similarly. At this point, all methods nearly always include all signal variables in their entertained sets.
		\item \Protolasso\ entertained proportion performance increases with the number of selected clusters (along horizontal axis). Remember that the initial clustering imposes some restrictions on which variables make it into the entertained set. If we select only one cluster, but signals are distributed over more than one cluster, many signal variables do not make it into the entertained set. Since we allow the original lasso to select as many variables as there are in the entertained set of the \protolasso\, without regard to a clustering, it is free to select signal variables from other clusters not currently considered by the \protolasso\ procedure.
		\item The more clusters in the original clustering; the more potential prototypes. If we choose the number of clusters in excess of the true number of groups, we allow potentially correlated variables into the second stage lasso, leading to the same variable selection problems discussed before.
		\item Results are qualitatively similar for the single block dataset. We omit these plots in the interest of saving space.
	\end{itemize}
	
	Overall then, it would seem that \protolasso\ includes the appropriate variables in its entertained set. The slight relaxation implied by the definition of the entertained set allows us to overcome the erratic variable selection issue encountered by the lasso amidst high correlations.	

	\subsection{Confidence interval width}
	The post selection inference framework provides a simple way to make such inferences on a surprising variety of selection procedures. Confidence intervals are constructed for the selected signals and their widths depend, inter alia, on the nature of the selection procedure via the $A$ and $b$ matrices characteristic of the procedure. One can ask how interval widths compare for different procedures. Although the three procedures do not necessarily (rarely, in fact) compute confidence intervals for the same selected set -- or even have the same target for the selected variable they do share -- we would still prefer a procedure that produces narrower intervals. The post selection framework provides guarantees on coverage, so we do not consider this here. We focus solely on relative interval widths, via the performance measure described in the previous subsection.
	
	\begin{figure}[hbtp]
		\centering
		\includegraphics[width=90mm]{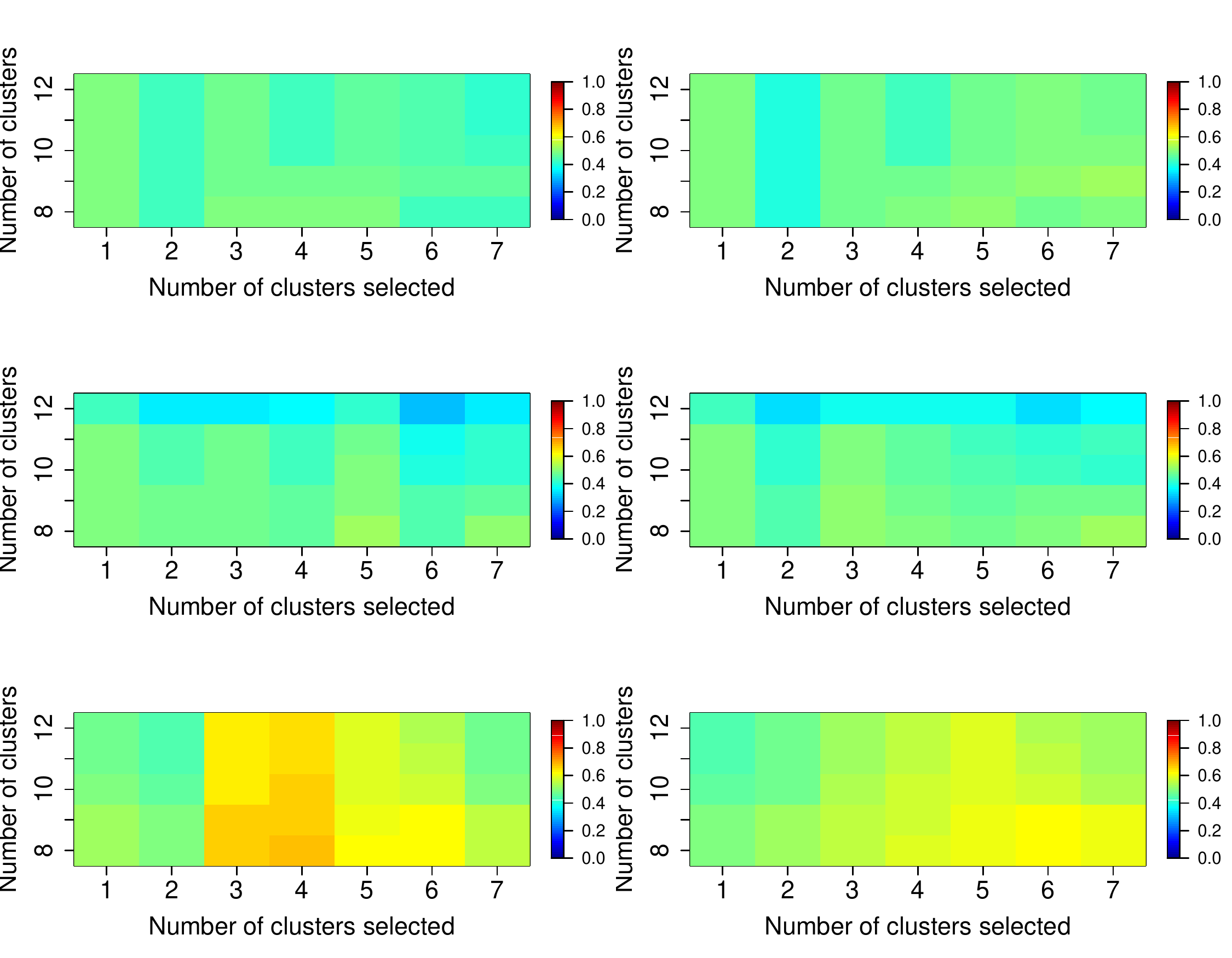}
		\caption{\emph{Heatmap of relative confidence width performance measures as function of the initially extracted clusters (vertical axis) and those selected by the second stage lasso (horizontal axis). Dataset is the block diagonal $X$ with the paired signal $\beta$ configuration. $\beta_* = 2$. Same setup as Figure~\ref{fig:ep_heat1}.}}
		\label{fig:ci_heat1}
	\end{figure}
	
	\begin{figure}[hbtp]
		\centering
		\includegraphics[width=90mm]{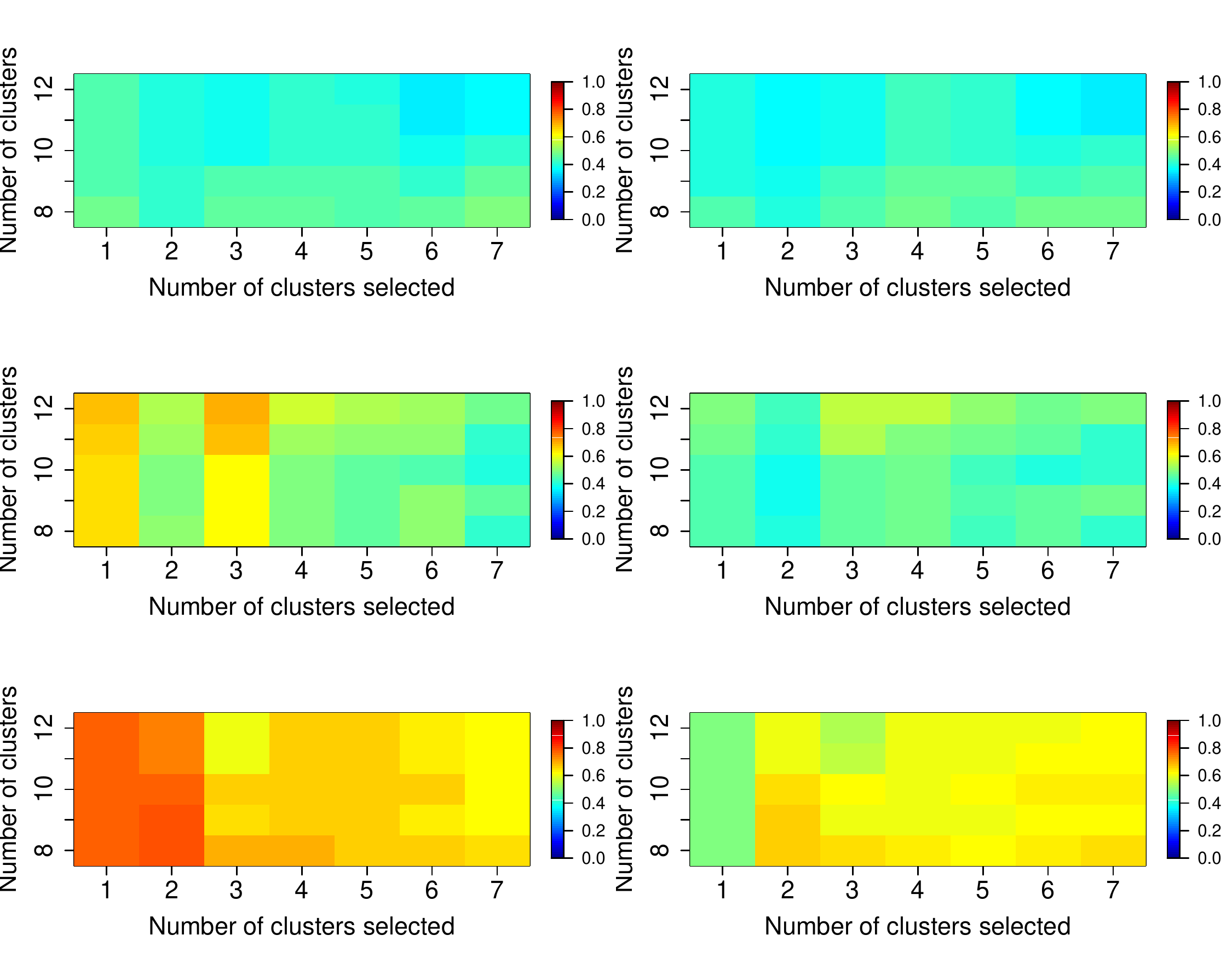}
		\caption{\emph{Heatmap of relative confidence width performance measures as function of the initially extracted clusters (vertical axis) and those selected by the second stage lasso (horizontal axis). Dataset is the single block $X$ with the tight signal $\beta$ configuration. $\beta_* = 2$. Same setup as Figure~\ref{fig:ep_heat1}.}}
		\label{fig:ci_heat2}
	\end{figure}
	
	Figures~\ref{fig:ci_heat1} and~\ref{fig:ci_heat2} show heat maps of the relative confidence interval width performance of the \protolasso\ procedure against the other two methods. Figure~\ref{fig:ci_heat1} uses a block diagonal matrix, while Figure~\ref{fig:ci_heat2} uses a single block matrix. The signal size is large ($\beta_* = 2$) so as to eliminate variable selection issues (all methods usually select all signals into their entertained sets) and focus only on the subsequent confidence interval widths. Some observations:
	
	\begin{itemize}
		\item The \protolasso\ rarely constructs confidence intervals systematically much wider than the other two methods. Its worst performance occurs in the low correlation ($\rho = 0.1$) setting when we overselect the number of clusters. \Protolasso\ seems to require a reasonably good estimate of the grouping structure for it to generate narrower intervals.
		\item \Protolasso's relative performance improves markedly as $\rho$ increases. At $\rho = 0.7$ (bottom rows, both figures), we see that intervals are in general shorter than both competitors, especially the lasso, in particular when we select roughly the correct number of signal clusters (vertical bands around 3/4 selected clusters in Figure~\ref{fig:ci_heat1} and 1/2 in Figure~\ref{fig:ci_heat2}).
		\item Note that Figure~\ref{fig:ci_heat2} has the tight distribution of the signal, all on variables correlated with each other. The lasso might select all the variables here, given the signal size. However, the high correlation amongst them (say when $\rho = 0.7$) causes some numerical instability in the constraint matrices $A^{lasso}$ and $b^{lasso}$. This leads to wider intervals. \Protolasso\ does not suffer from this. It selects only one of these variables (as they are all presumably in the same cluster) and its constraint matrices are not affected by the large correlation. However, all signals are still in the entertained set and we can still compute confidence intervals for them.
	\end{itemize}
	
	Overall then,  \protolasso\ performs admirably in the face of increased correlation, but does comparatively less well in low correlation settings where we estimate too many clusters in the original clustering. A similar pattern to the entertained proportion performance then.

\section{Marginal testing:  Prototest}
\label{sec:prototest}
In the previous sections, we  focussed on the  regression problem, applying the lasso to the selected prototypes from clustering of the features, a procedure we called \Protolasso.
Here we briefly discuss the simpler problem of marginal testing.

The first two steps  in this method are the same as in \Protolasso:  we cluster the features, and choose a representative prototype from each cluster, being the one most correlated with the outcome.
But instead of fitting the lasso to these prototypes, we carry out marginal  hypothesis testing of each one.  This inference is based on the selection-adjusted p-values
using (\ref{eqn:A1}), (\ref{eqn:b1}) to account for they choice or prototype.  We do not include  (\ref{eqn:A2}), (\ref{eqn:b2}), since the lasso is not used.
We call the resulting method the \prototest\ procedure. P-values for non-prototypes can be derived in much the same way as we did in \protolasso.

\begin{figure}[hbtp]
		\centering
		\includegraphics[width=4in]{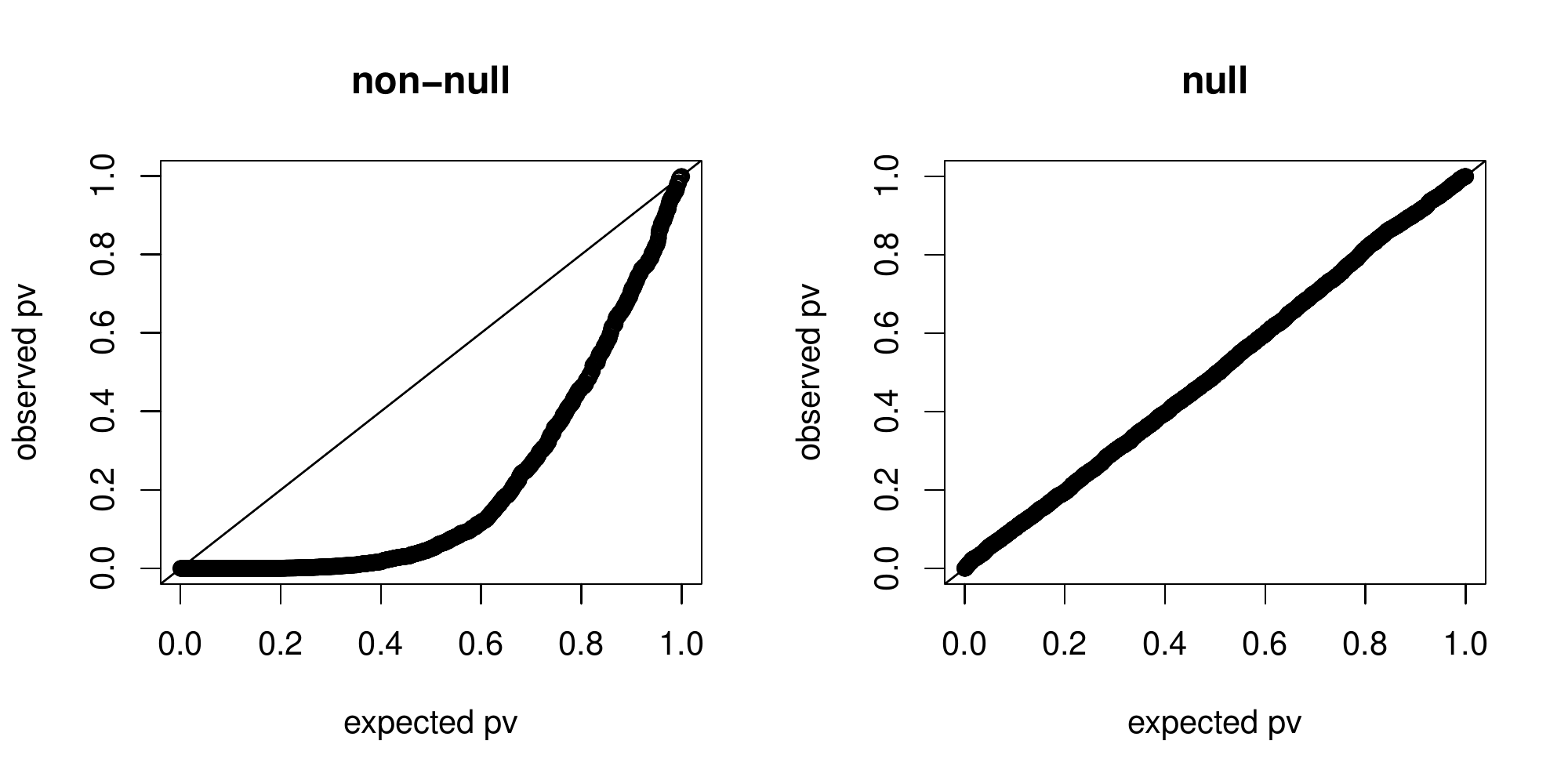}
			\caption[fig:testingpv]{\em \Prototest: Quantile-quantile plots of non-null and null p-values for simulated data example.}
				\label{fig:testingpv}
	\end{figure}

Figure \ref{fig:testingpv} shows an example. We generated data from a standard Gaussian linear model: there were 10 groups of 6 features each, having within-group pairwise correlation of 0.7;  3 of the groups having a single  signal feature.
We orthogonalized the features of the other 7 group with respect to the 3 groups.
%illustrate  asymptotic FDR control via BH; could also use permutations.
The p-values over 200 simulations for the non-null and null groups are shown in the Figure. We see that the non-null p-values exhibit reasonable  power while the null p-values
are very close to uniform, as predicted by our  theory.

Note that we have implicitly considered ``non-null' any cluster containing a signal feature, even if the prototype chosen for  that cluster is not the signal feature.
This seems reasonable, since the features in each cluster are highly correlated.

The basic definitions of a false positive call and the false discovery rate, become murky when the items (feature) under consideration are correlated.
This very issue is the central topic, for example,  of the paper by \citet{GHT2014}.  The issue is as follows. In a regression setting for example,
suppose that two predictors $X_1$ and $X_2$ are very highly correlated but only $X_1$ has a non-zero population coefficient.
If we apply a section procedure and it chooses $X_2$, is this a false positive selection? With standard definitions of false positive and FDR, it would be.
But intuitively, it seems like it should be considered a true positive.
\citet{GHT2014} propose an alternate definition of false positive rate and FDR to address this issue, what they denote the UVR (uninformative variable rate).
We will not discuss the details here.  But we note that by considering correlated clusters as non-null or null, we skirt this issue.

\begin{figure}[hbtp]
		\centering
		\includegraphics[width=4in]{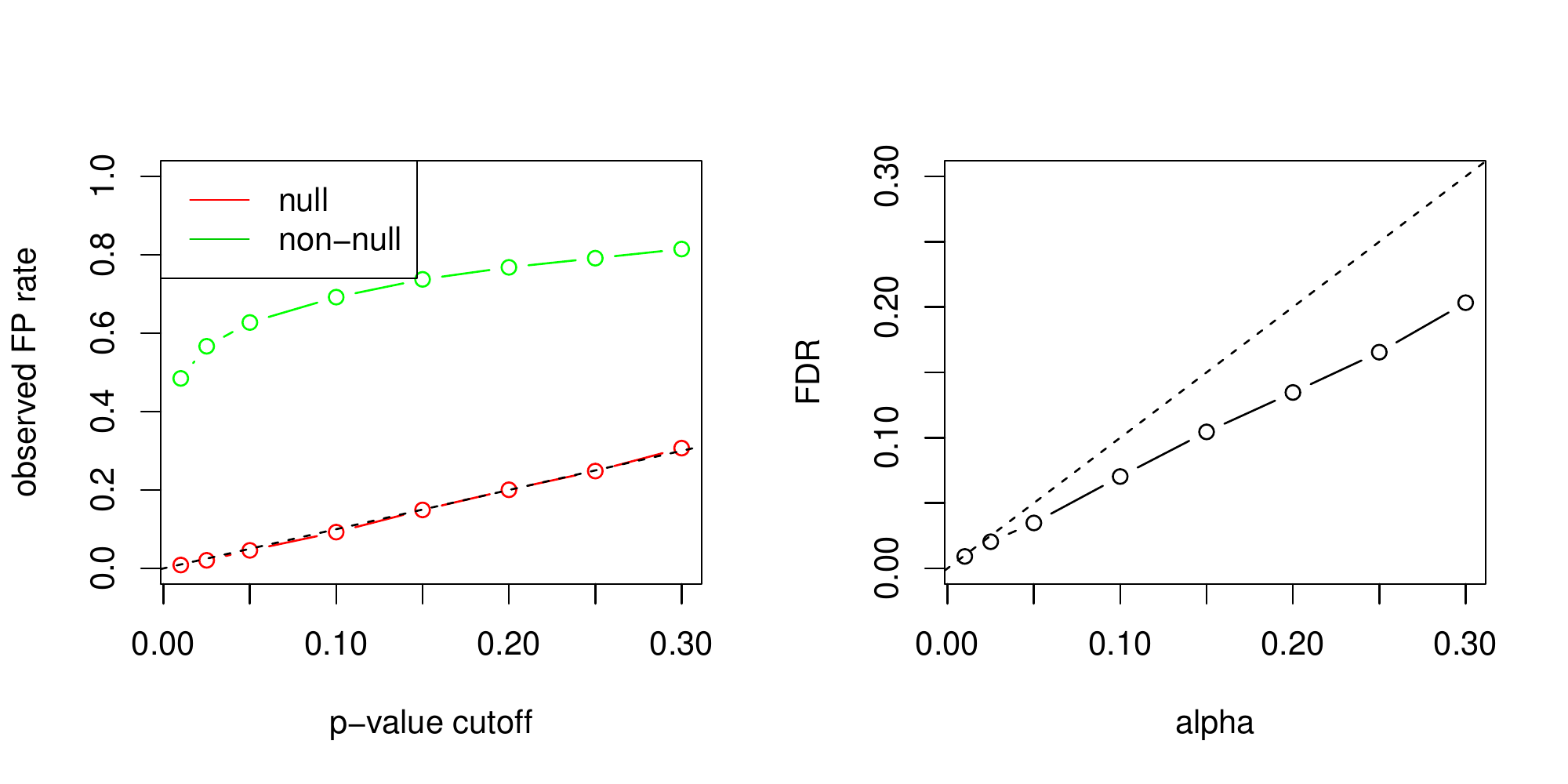}
				\caption[fig:testingpv]{\em For the same simulation setup of Figure \ref{fig:testingpv},  the left panel
		shows that the  false positive rates for the null clusters follow their expectation closely. 
In the  right panel the standard Benjamini-Hochberg  FDR controlllng procedure has been applied, for various nominal FDR  values $\alpha$.
The achieved FDR shown and lies below the $45^o$ line (broken), as it should.}
\label{fig:testing}
\end{figure}

Using this idea, we computed the false positive rates for our simulated sample in the left panel of Figure \ref{fig:testing}.
We see that the false positive rates for the null clusters following their expectation closely. 
In the  right panel we applied standard Benjamini-Hochberg  FDR control to these p-values .
We see that the FDR is indeed controlled at the nominal level. Since the p-values here are not exactly independent, the BH finite-sample control does not follow. However the asymptotic FDR control results of
\citet{STS2005} will likely be applicable.

In many testing situations, especially in computational biology,  the outcome is $y$ is binary, for example case-control status.
Then the Gaussian linear model is not appropriate. In that setting we can extend the polyhedral lemma and \prototest\
to yield asymptotically valid selection-adjusted p-values. We give some details next.

Consider a likelihood-based regression model with natural parameter 
$\eta=X\beta$ and log-likelihood $\ell(\beta)$. Let
\smash{$\hat{U}(\beta)$} and \smash{$\hat\sI(\beta)$} be the sample
score vector and information matrix, and $U(\beta)$ and
$\sI(\beta)$ be the population versions.  Assuming the existence of
some true coefficient vector $\beta^*$, under standard asymptotics
with $p$ fixed and $n \rightarrow \infty$, we have
\begin{equation}
\hat U(\beta^*) \sim 
N\big(0, \sI(\beta^*)\big).
\end{equation}

In finite samples, we can use a normal approximation to the score.
The usual score test enters the variable $j_1$ with sign $s_1$ that
maximizes \smash{$s \cdot\hat{U}_j(0)/\hat\sI_j(0)^{1/2}$} over all
variables $j$ and signs $s$.  Suppose that wish to test the hypothesis 
$\beta^*_{j_1}=0$, by testing that $U_{j_1}(0)=0$.  To do this, we use
the approximation 
\begin{equation}
\label{eq:approx1}
\hat{U}(0) \sim N\big(0, \hat\sI(0)\big), 
\end{equation}
and apply the polyhedral selection and inference lemmas, with 
constraints of the form 
\begin{equation}
s_1 c_{j_1}^T  \hat{U}_j (0)\geq \pm c_j^T \hat{U}_j(0) \;\;\; 
{\rm for}\; {\rm all } \;j \neq j_1
\end{equation}
Above, $c_j$ contains \smash{$\hat\sI_j(0)^{-1/2}$} in the $j$th
component and is 0 in all other components. 
The prototype selection in \prototest\ would choose the feature with
the largest value of the score test, and the above allows us to compute selection-adjusted p-values.

\begin{figure}
  \centering
  \includegraphics[width=3.25in]{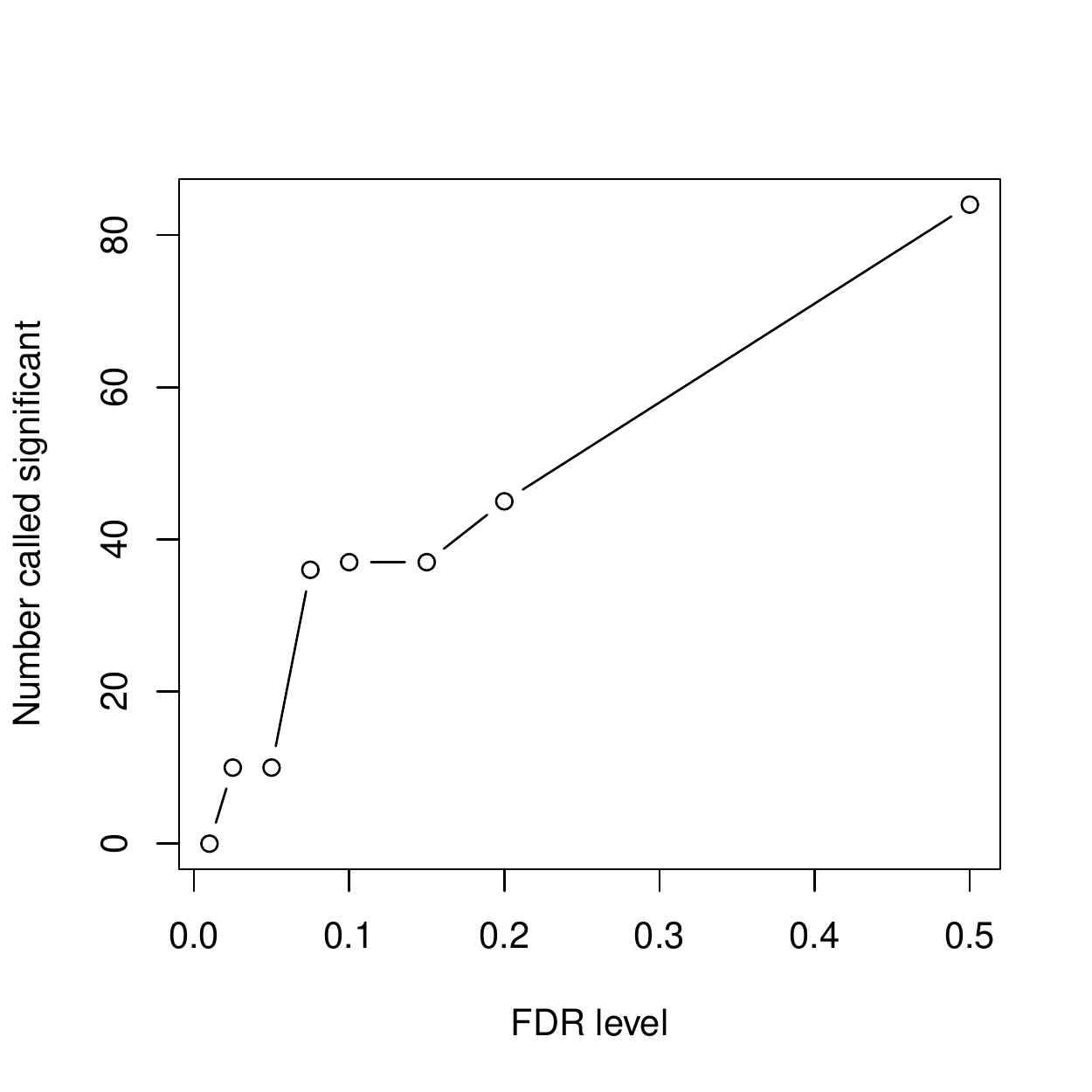}
  \caption[fig:p53fdr]{\it  p53 data: number of significant gene sets found at various FDR levels.}
\label{fig:p53fdr}
\end{figure}

Next we consider the microarray example from \citet{sub2005} and
\citet{ET2007}:  there are $p=10,100$ genes
and $N=50$ arrays, relating to cell lines with normal or mutated
states for the p53 factor, $n_1=17$ normal  and $n_2=33$ mutated.One
We coded this as $y=0,1$.
\citet{sub2005} curated several collections of genesets,
representing organization of genes into functional categories. Here we
consider the ``C2'' grouping into
 522 overlapping gene
sets, representing cell pathways. As in \cite{ET2007},  we restrict attention to the 395 sets having $\geq 10$ members.
The approaches of  \citet{sub2005} and \citet{ET2007} summarize
the genesets with some overall score, and then use a permutation
test to assess the overall significance. Here we apply {\tt
  prototest}, choosing the most representative gene from each gene
set, and then compute exact p-values
to account for the selection.
 Figure \ref{fig:p53fdr} shows the results of applying the BH rule
at various FDR levels.
Qualitatively the number of  significant sets is in line with the analysis found in \citet{ET2007}.
Due to space constraints,  we defer a more in-depth study of \prototest\  for the logistic and other generalized linear models to a future paper.
\section{FDR control for the \protolasso\ procedure} 
\label{sec:fdr}

In this section we adapt the {\em knockoff} framework of \cite{BC2014}  to the \protolasso\ method, in order to  obtain False Discovery Rate (FDR) control.
	
	FDR was first introduced in \citet{BHfdr} in a multiple testing framework. The authors define the FDR:
	\[
		E\left[\frac{V}{\max\{R, 1\}}\right]
	\]
	where $R$ is the number of the (multiple) null hypotheses rejected and $V$ the number of those rejected incorrectly. The introduction of this controlling measure -- and their procedure guaranteed to control it below some predetermined proportion $q$ -- changed the manner in which practitioners considered multiple testing. 
	
	The seminal paper has spawned much subsequent research. One paper of particular interest is that of \citet{BC2014}, which introduces a variable selection technique guaranteed to control FDR. In a variable selection setting, one would associate $R$ with the number of variables selected and $V$ with the number of null variables selected by the procedure. Their procedure is called \textit{knockoff screening} and proceeds in a sequence of steps:
	\begin{itemize}
		\item \textbf{Formation of the ``knockoff" matrix} $\tilde{X}$ of the predictor matrix $X$. This matrix is constructed so that $X^\top X = \tilde{X}^\top\tilde{X}$ and $X^\top\tilde{X} = X^\top X - {\rm diag}(s)$. So the knockoff matrix preserves all the correlations of the original matrix, both amongst its own variables and those of the original variables, but reduces the correlation between knockoffs and their original counterparts. The authors describe how to construct $\tilde{X}$ and propose how to select the $0 \leq s_j \leq 1$.
		\item \textbf{Construction of statistics for each pair of original and knockoff variables.} Each pair gets its own statistic $W_j$, $j = 1, 2, \dots, p$. If $W_j$ is large and positive, then there is evidence against the hypothesis that this variable is null.
		\item \textbf{Definition of data dependent threshold}, $T = \min\left[t \in \mathcal{W}: \frac{1 + \#\{j: W_j \leq -t\}}{\#\{j: W_j \geq t\} \vee 1 } \leq q\right]$, where $\mathcal{W} = \{|W_j|: j = 1, \dots, p\} \setminus \{0\}$.
		\item \textbf{Select features} with $W_j \geq T$ to control FDR at $q$.
	\end{itemize}
	
	The shrewdness of the method is in the manner in which they construct the knockoffs and the statistics $W_j$. They present many examples of potential $W_j$ constructions, but note that their FDR result holds as long as the $W_j$ obey two properties:
	\begin{enumerate}
		\item \textbf{Sufficiency property}: $W_j$ should depend on the data only via the (column concatenated) Gram matrix $\left[\begin{array}{c c} X &\tilde{X}\end{array}\right]^\top \left[\begin{array}{c c} X &\tilde{X}\end{array}\right]$ and feature-response inner products $\left[\begin{array}{c c} X &\tilde{X}\end{array}\right]^\top y$.
		\item \textbf{Antisymmetry property}: detailed in the reference.
	\end{enumerate}
	
	One example of valid $W_j$, mentioned in the paper, and used by us later in the section, is to fit a lasso regression of $y$ on $\left[\begin{array}{c c} X &\tilde{X}\end{array}\right]$, storing the largest value of the regularization parameter $\lambda$ such that the variable enters the model, i.e.
	\[
		Z_j = {\rm sup}\{\lambda: \hat{\beta}_j(\lambda) \neq 0\}
	\]
	with the knockoff equivalent $\tilde{Z}_j$ similarly defined. Setting $W_j = Z_j - \tilde{Z}_j$ achieves the desired properties.

	Proof of their results follows from a super-martingale argument which flows quite elegantly, considering the obviously complicated distributional properties of the $W_j$ and $T$. In fact, the crux of the result hinges on a succession of lemmas (numbers 1, 2 and 3 in the reference) that establish certain exchangeability results, making the rest of the analysis essentially independent of the underlying distribution of the $W_j$. 
	
	In the remainder of this section, we describe a method modifying the knockoff procedure to operate at the prototype level, meshing it nicely with the initial clustering-prototyping step championed in this paper. We design the procedure to allow replication of Lemmas 1, 2 and 3 of \citet{BC2014}, hence establishing FDR control for our procedure at the prototype level. The section concludes with a brief experiment demonstrating an application of our knockoff method to \Protolasso.

	\subsection{A Knockoff procedure for \protolasso}
	\label{sec:protoknockoff}
	One wonders whether the prototyping-clustering and knockoff procedures can be seamlessly interleaved. Suppose we have column centered and standardized the predictor matrix $X$. Recognizing that the construction of knockoff $\tilde{X}$ ensures $X^\top X = \tilde{X}^\top\tilde{X}$, one realizes that any clustering based on the correlation metric detailed above produces the \textit{same} clustering on both sets of columns. An initial instinct is to cluster the columns of $X$ (and, by extension, for $\tilde{X}$) and then finding maximal marginal correlation prototypes for each separately, delivering prototypes $P = \{P_1, P_2, \dots P_K\}$ for the $K$ clusters of the columns of $X$ and similarly $\tilde{P} = \{\tilde{P}_1, \tilde{P}_2, \dots, \tilde{P}_K\}$ for $\tilde{X}$. Note that the separate computation of the two sets of prototypes does not guarantee that $P_j = \tilde{P}_j$, despite the initial clustering being the same. This is because the maximal marginal correlation column (with response $y$) in a given cluster need not be the same in $X$ as it is in $\tilde{X}$. The knockoff procedure alters correlations with the response. One would then venture to proceed by forming matrices $X_P$ and $\tilde{X}_{\tilde{P}}$ and performing the knockoff procedure as described by \citet{BC2014} on the response $y$ with augmented matrix $\left[\begin{array}{c c} X_P &\tilde{X}_{\tilde{P}}\end{array}\right]$.
%%\marginpar{I removed the name proto-knockoff--  seemed to be overdoing it!}	
	Although we believe this procedure \textit{could} control FDR, we found in experiments that it has very low power (ability to detect prototypes in clusters containing signal variables). We suspect that this has to do with the separate selection of prototypes in the original and knockoff matrices, especially in matrices where we encounter high correlation. In high correlations, prototypes are less likely to be the actual signal variables from a cluster (since every variable is a good surrogate for all the others and non-signal variables may present as having high correlation with the response). Also, we choose the most correlated knockoff variable from the knockoff cluster, which reduces the apparent strength of the original variable in the subsequent computation of $W_j$ (even if this original variable comes from a signal cluster). We see fewer large positive $W_j$ and knockoff screening tends not to select any variables. In sum: prototypes have sight of $y$ before subsequent analysis and selected prototypes tend to mismatch over original and prototype matrices: $P_j \neq \tilde{P}_j$. This reduces power.
	
\begin{figure}[htb]
		\centering
		\includegraphics[width=120mm] {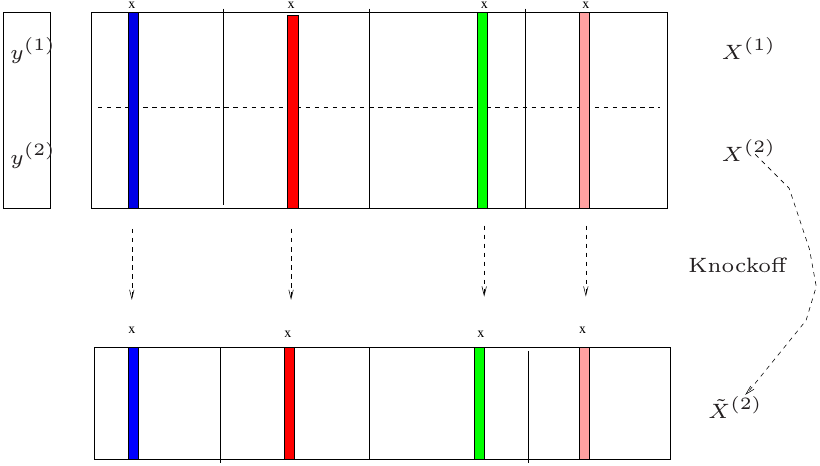}
		\caption[fig:knockoff]{\em Schematic of our knockoff strategy with sample splitting:  Columns of original matrix are clustered using all of $X$. Solid vertical black lines delimit clusters. There are four clusters in the figure. Notice how the clusters are translated exactly to the knockoff matrix. Original matrix is then split into $X^{(1)}$ and $X^{(2)}$ and response into $y^{(1)}$ and $y^{(2)}$ (horizontal dotted line in top part of the figure). Prototypes are found for the clusters using only $X^{(1)}$ and $y^{(1)}$. These prototypes are highlighted by coloured vertical bars. Notice how they too are translated directly over to the knockoff matrix. Knockoff matrix $\tilde{X}^{(2)}$ is obtained from the \textit{whole} $X^{(2)}$ before reducing to prototypes. Notice that a knockoff of $X^{(1)}$ is not created.}
		\label{fig:knockoff}
	\end{figure}

	Our  procedure addresses each of these shortcomings and orders steps in such a way to ensure the exchangeability lemmas of \citet{BC2014} still hold. The procedure is illustrated in Figure~\ref{fig:knockoff} and detailed as follows:
	\vskip .25in
	\centerline{\bf Knockoff procedure for \protolasso}
	\begin{enumerate}
		\item \textbf{Cluster columns} of $X$ and select number of clusters $K$, producing clustering $\{C_1, C_2, \dots, C_K\}$.
		\item \textbf{Split the data} by rows into two (roughly) equal parts: $y = \left(\begin{array}{c} y^{(1)} \\ y^{(2)} \end{array}\right)$ and $X = \left(\begin{array}{c} X^{(1)} \\ X^{(2)} \end{array}\right)$.
		\item \textbf{Prototype clusters} via maximal marginal correlations, as before, using only $y^{(1)}$ and $X^{(1)}$. This yields prototype set $\hat{P} = \{\hat{P}_1, \hat{P}_2, \dots, \hat{P}_K\}$. $y^{(1)}$ and $X^{(1)}$ are now excluded from further analysis.
		\item \textbf{Form knockoff matrix} $\tilde{X}^{(2)}$ from $X^{(2)}$ as described in \citet{BC2014}. It is essential that we form the knockoff matrix here using all columns of $X^{(2)}$. This is to ensure that the exchangeability results hold for our procedure. Details are in Appendix B.
		\item \textbf{Reduce to matrices} $X^{(2)}_{\hat{P}}$ and $\tilde{X}^{(2)}_{\hat{P}}$ by selecting the relevant (and same) columns in $X^{(2)}$ and $\tilde{X}^{(2)}$ respectively.
		\item \textbf{Proceed with knockoff screening} using $y^{(2)}$ and $\left[\begin{array}{c c} X^{(2)}_{\hat{P}} & \tilde{X}^{(2)}_{\hat{P}}  \end{array}\right]$.
	\end{enumerate}
	\vskip .25in
	We show in the Appendix B how this procedure replicates the exchangeability lemmas in \citet{BC2014}. Hence our procedure exactly controls FDR in finite samples, as summarized in the following lemma:
	\begin{lemma}
		For any $q \in [0,1]$, let $W^{(2)}_j$, $j\in \hat{P}$ be the statistics testing the knockoff-original pairs in the knockoff procedure for \protolasso, where $\hat{P} = \{\hat{P_1}, \hat{P_2}, \dots, \hat{P_K}\}$ is the set of prototypes selected in the first stage of the \protolasso\ procedure. Furthermore, let
		\[
			T = \min\left[t \in \mathcal{W}: \frac{1 + \#\{j \in \hat{P}: W^{(2)}_j \leq -t\}}{\#\{j \in \hat{P}: W^{(2)}_j \geq t\} \vee 1 } \leq q\right]
		\]
		where $\mathcal{W} = \{|W^{(2)}_j|: j \in \hat{P} \} \setminus \{0\}$. Finally, let $\hat{S}_{\hat{P}} = \{j \in \hat{P}: W^{(2)}_j \geq T\}$. Then
		\[
			E\left[\frac{\#\{j : \beta_{\hat{P}_j} = 0 \,\, {\rm and} \,\, j \in \hat{S}_{\hat{P}}\}}{\#\{j: j \in \hat{S}_{\hat{P}}\} \vee 1}\right] \leq q
		\]
		where the expectation is taken over the distribution of $y^{(1)}$, which generates the prototype set $\hat{P}$, and that of $y^{(2)}$, which generates the statistics $W^{(2)}_j$.

	\end{lemma}
	
	 In the next section we perform and experiment to study the power of the procedure.
	
	\subsection{Simulation experiment}
	\label{sec:knockoff_sim}
	The predictor matrices $X^{(1)}$ and $X^{(2)}$ for the simulation experiment of this section are generated like the block diagonal type matrices generated throughout the paper. Each matrix is generated once, with $n = 200$ rows and $p = 100$ columns, comprising of 10 blocks of 10 predictors each, with the pairwise correlation $\rho = 0.5$ in each block.
	
	We consider three configurations for $\beta$. Most of the elements are set to zero. Each configuration has five clusters with non-zero coefficients, but different numbers of non-zero coefficients within a cluster. The three configurations are:
	\begin{enumerate}
		\item Coefficients at indices $j = 1, 11, 21, 31, 41$ are set to $\beta_*$. Here we have one signal variable per cluster.
		\item Coefficients at indices $j = 1, 2, 11, 12, 21, 22, 31, 32, 41, 42$ are set to $\beta_*$. So, two signal variables per cluster.
		\item Coefficients at indices $j = 1, 2, 3, 11, 12, 13, 21, 22, 23, 31, 32, 33, 41, 42, 43$ are set to $\beta_*$. Three signals per cluster.
	\end{enumerate}
	Different numbers of signal variables in clusters are tried, because considerations from Appendix B suggest that having multiple signals, but only one prototype from a cluster, could challenge FDR control unless we construct the knockoff matrix $\tilde{X}^{(2)}$ correctly. Our results seem to suggest we do have FDR control. Signal amplitude is varied, setting $\beta_* = 1, 2, \dots, 9$.
	
\begin{figure}[htb]
		\centering
		\includegraphics[width=110mm]{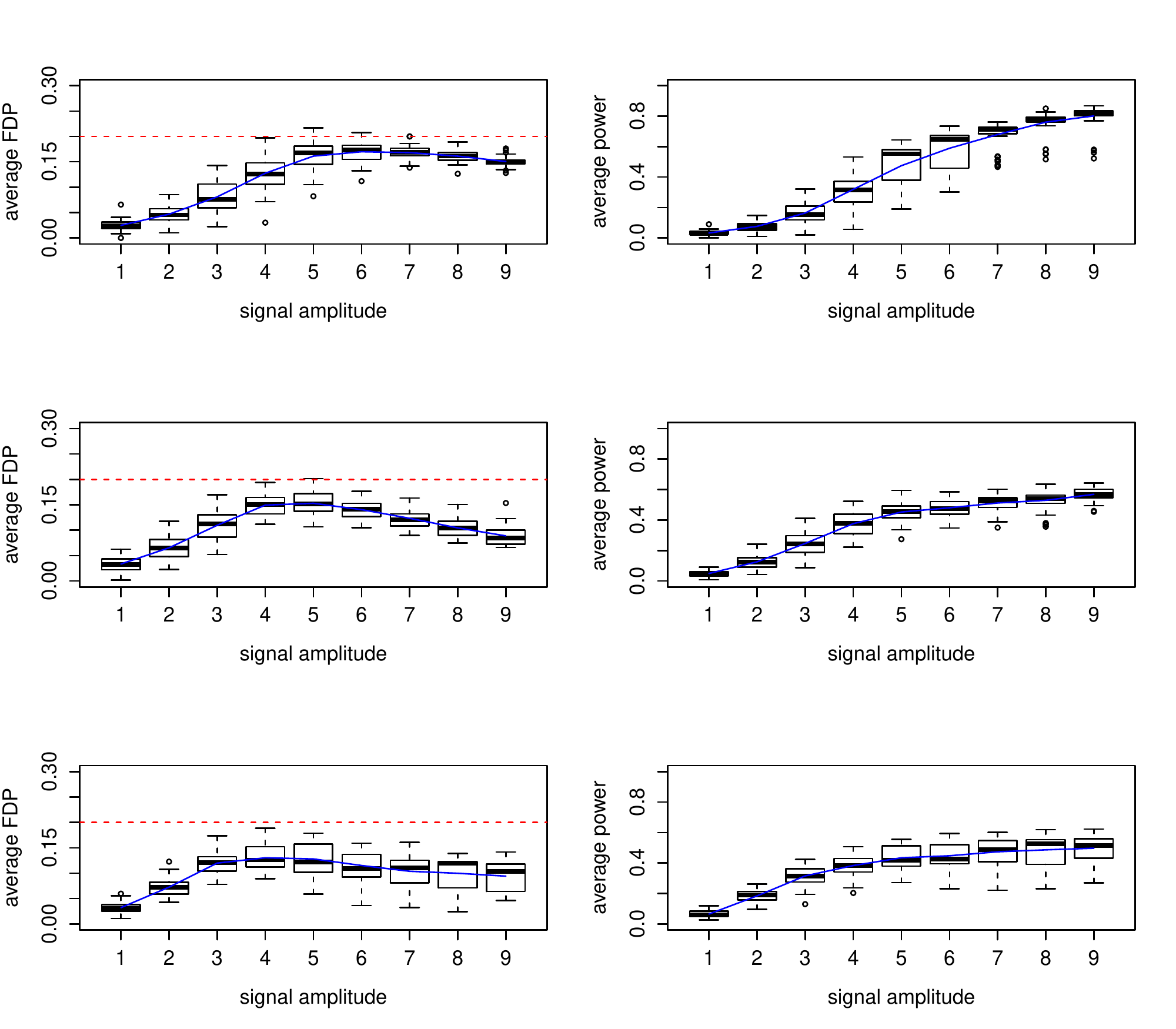}
		\caption{\emph{Boxplots of average FDP (\textbf{left panels}) and average power (\textbf{right panels}) over $S_1 = 50$ different prototype set realizations for the knockoff procedure for \protolasso\ (means in blue). Plotted as function of signal amplitude $\beta_*$. Data is a block diagonal type matrix with $n = 200$ rows and $p = 100$ columns: 10 blocks of 10 variables each, $\rho = 0.5$. \textbf{Top row:} Single signal per cluster. \textbf{Middle row:} Two signals per cluster. \textbf{Bottom row:} Three signals per cluster. Red dotted horizontal lines drawn at FDR control level $q = 0.2$.}}
		\label{fig:fdp_power}
	\end{figure}

	At each setting for the coefficient vector (there are $3 \times 9 = 27$), we generate $S_1 = 50$ replications of $y^{(1)}$ and use it in conjunction with $X^{(1)}$ to find prototypes for the $K = 10$ clusters. At each of these $S_1 = 50$ realizations of the prototype set $\hat{P}$, we run $S_2 = 100$ replications of the knockoff procedure for \protolasso, computing the false discovery proportion (FDP) amongst the prototypes and the power -- the proportion of prototypes selected that occur in a cluster with \textit{at least one signal variable}. These two quantities are averaged over the $S_2 = 100$ replications to provide estimates of FDR and average power at a given prototype set. FDR is controlled at $q = 0.2$. Figure~\ref{fig:fdp_power} shows boxplots of the estimated FDR and average power over the $S_1 = 50$ prototype set realizations, as function of $\beta_*$.
	
	Notice that the blue lines (our overall of FDR) are always well below the red FDR control line at $q = 0.2$. This despite one or two replications in configurations coming in above the threshold. This is probably due to variation in estimating the FDR at a given prototype realization. Average power increases with signal amplitude, but does so more slowly the more signal variables we put into clusters. This is probably since our prototyping procedure screens more signals at the prototyping stage, with the second stage lasso procedure (where we compute the $W_j$) involving variables with proportionately small amounts of signal. It might be more difficult for the true signal's $Z_j$ to come to the fore above their knockoff counterparts $\tilde{Z}_j$, leading to fewer true signal detections. Overall then, we seem to have FDR control, with reasonable signal detection power.
	
	\section{Real datasets}
	In this section we apply the \protolasso\ procedure to two real datasets, demonstrating example applications of the procedure. The two datasets are the data used in \citet{conlondata} and some data in HIV mutations. Neither dataset is wide $p > n$. This is not a problem for our method. It applies equally to datasets with $n > p$. 
	
	\subsection{Conlon et. al.  data}
	
\begin{figure}[htb]
		\centering
		\includegraphics[width=120mm]{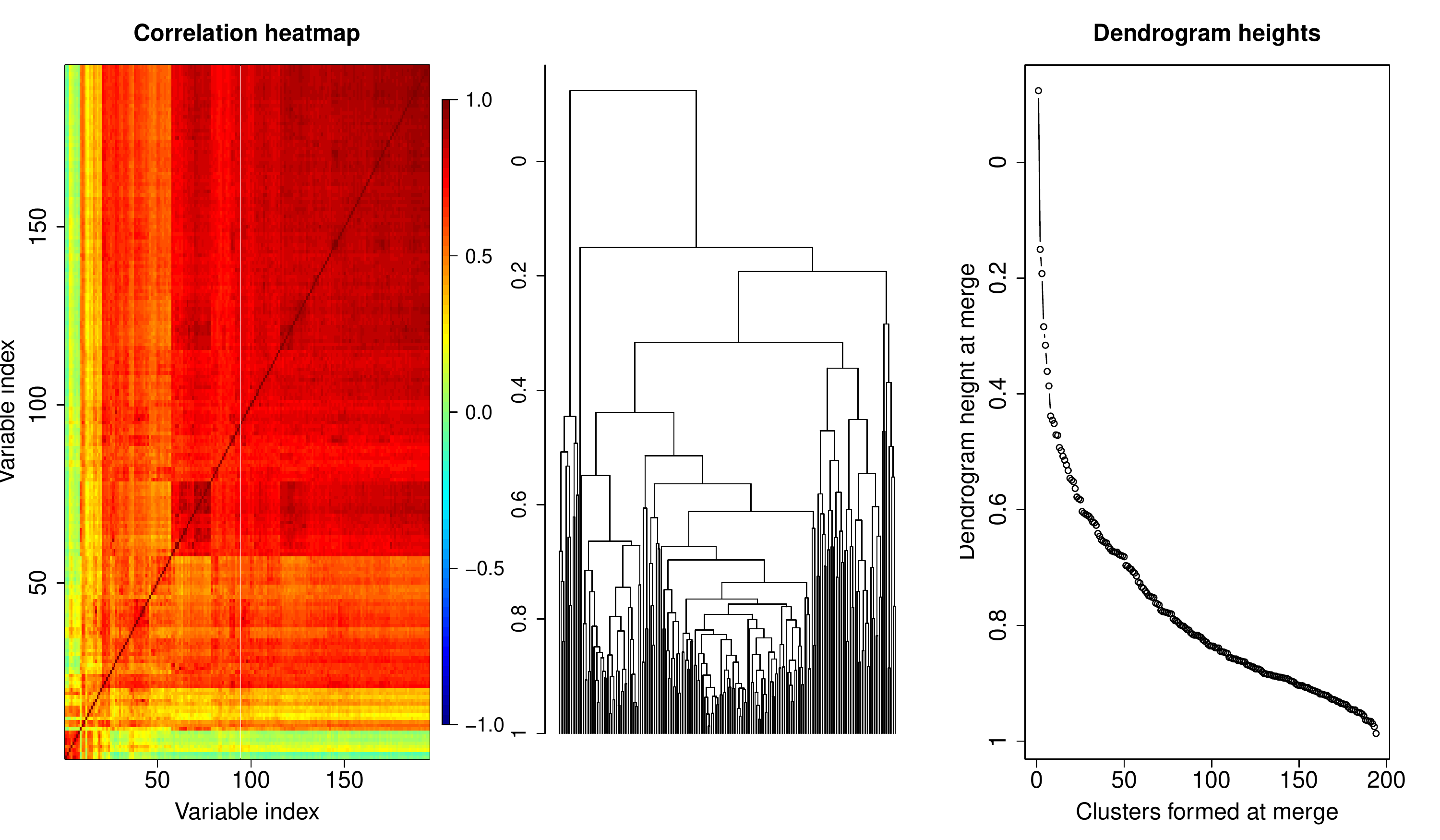}
		\caption{\emph{\textbf{Left panel:} Heatmap of pairwise correlations amongst (rearranged) columns of \citet{conlondata} data. Note the absence of any distinct clustering structure. Most variables are quite highly correlated with one another. \textbf{Center panel:} Dendrogram of hierarchical clustering with complete linkage based on one minus the correlation matrix. \textbf{Right panel:} Plot of dendrogram merge heights against the number of clusters formed at that merge. The panel, combine with the adjacent dendrogram seems to suggest that there are very few clusters in the data.}}
		\label{fig:conlon_heat}
	\end{figure}

	Figure~\ref{fig:conlon_heat} plots the initial analysis of the \citet{conlondata}. The data, initially used to demonstrate motif regression, has $n = 287$ rows and $p = 195$ columns, again mean centered and standardized (and permuted to get the correlated ones close together). Each column $j$ represents a candidate word or `motif', with $x_{ij}$ measuring how well a motif $j$ is represented in a DNA sequence $i$. The response $y_i$ is protein binding intensity in DNA sequence $i$.
	
	Again, we apply the gap statistic to estimate the number of clusters. This gives $K = 2$, which is not wholly surprising considering the nature of correlation heat map in Figure~\ref{fig:conlon_heat} -- notice the high correlation amongst all variables and little obvious grouping structure. Error variance is estimated once again via the least squares estimate $\hat{\sigma} = 4.8485$.
	
\begin{figure}[htb]
		\centering
		\includegraphics[width=120mm]{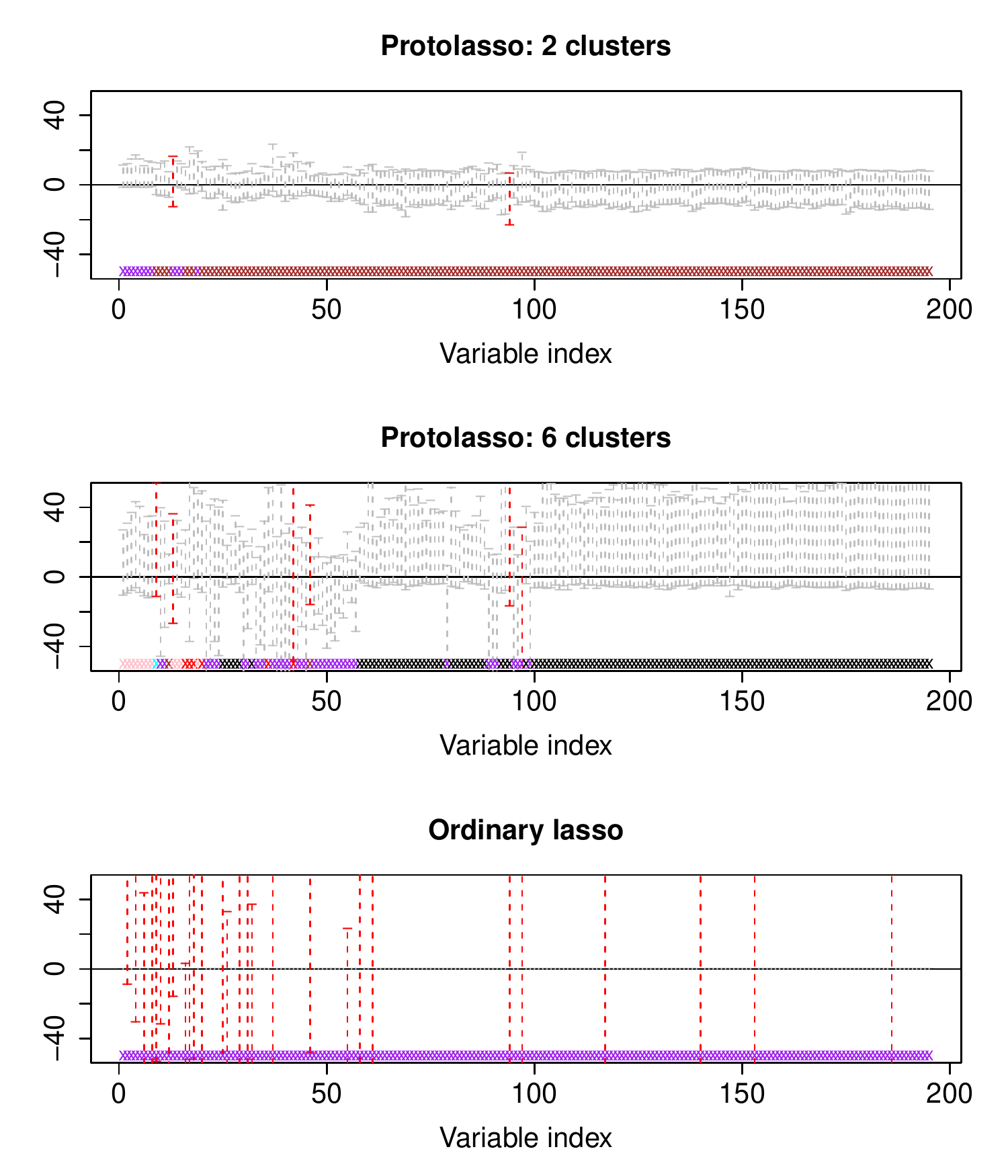}
		\caption{\emph{\textbf{Top panel:} $95\%$ post selection confidence intervals for variables in the entertained set of the \protolasso\ procedure applied to the \citet{conlondata} data. Selected prototypes have red intervals; other members have grey intervals. Number of clusters in initial clustering set to $K = 2$, while regularization parameter in second stage lasso on prototypes selected via 10-fold cross validation. Clustering structure represented by coloured crosses close to horizontal axis. \textbf{Center panel:} Same as top panel, but with $K = 6$. \textbf{Bottom panel:} $95\%$ post selection confidence intervals for variables selected by the ordinary lasso applied to the \citet{conlondata} data. Regularization parameter selected using 10-fold cross validation.}}
		\label{fig:conlon_ci}
	\end{figure}

	Figure~\ref{fig:conlon_ci} shows the CIs computed from the \protolasso\ (using both the gap statistic estimated $K = 2$ and then again with $K = 6$) and the ordinary lasso procedures. The figure demonstrates how correlation can hamper inference. Notice how wide all the intervals are for the ordinary lasso. This undoubtedly has to do with the numerical instability inherent in the inverses computed for the lasso post selection constraint matrices. For the \protolasso\ procedure with $K = 2$, we seem to select uncorrelated prototypes and the entertained set allows us to compute relatively narrow CIs for all variables, all of which include zero. When we have more prototypes in our second stage lasso ($K = 6$), the high correlation seems to widen our intervals again, but not as greatly as for the ordinary lasso.
	
	This dataset illustrates a good feature of the \protolasso\ procedure -- it selects only a few variables, breaking correlation where it can, leading to narrower CIs, but still allowing CIs to be computed for unselected variables. Still, it is not entirely immune to correlation if we specify an incorrect number of clusters. It should be noted that this dataset exhibits rather extreme correlation amongst its variables.
	 
		\subsection{NRTI data}
\begin{figure}[htb]
		\centering
		\includegraphics[width=120mm]{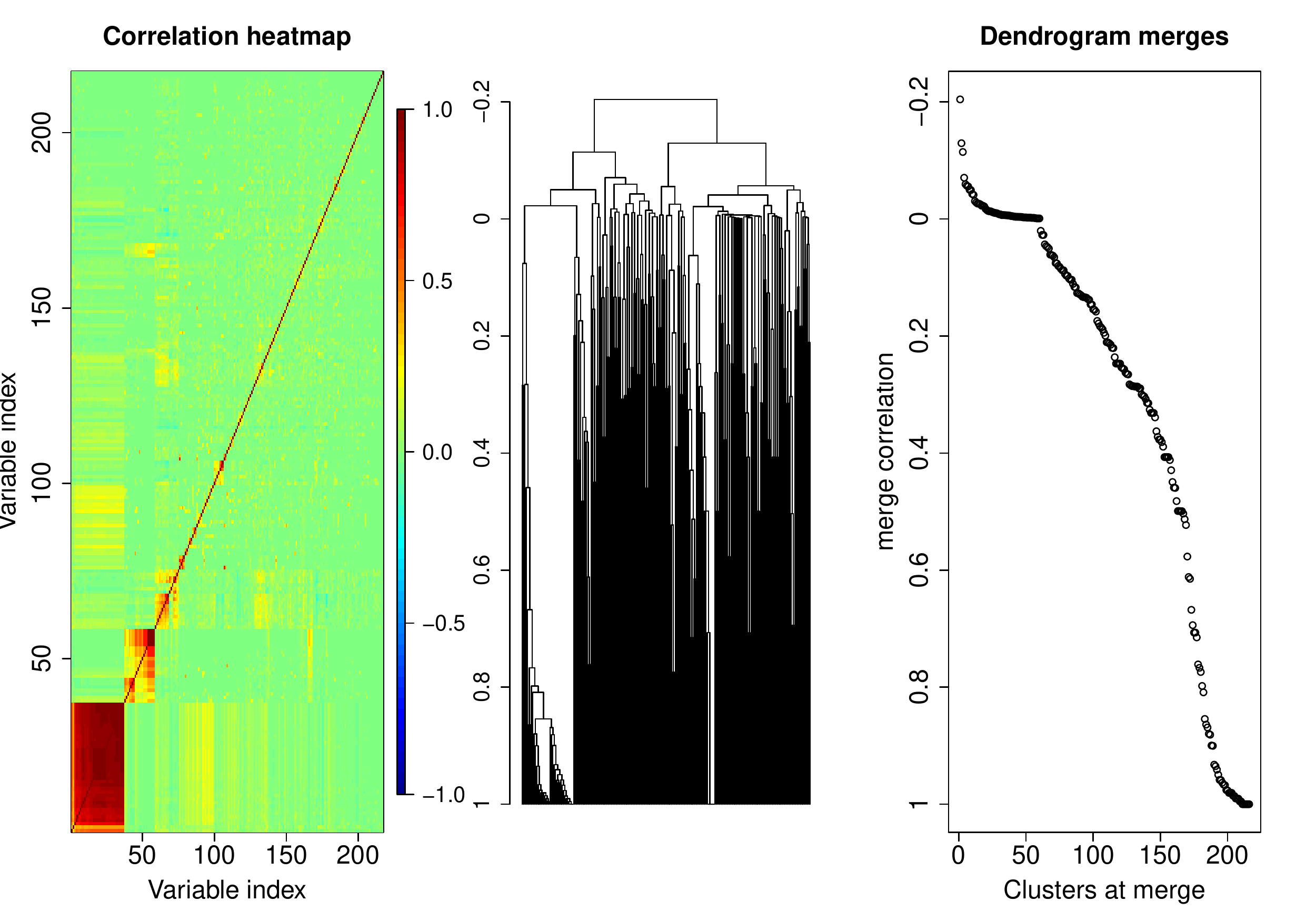}
		\caption{\emph{\textbf{Left panel:} Heatmap of pairwise correlations amongst (rearranged) columns of NRTI data. Note the one large, strong cluster, with two or three smaller, weaker clusters, with a smattering of pair clusters elsewhere. \textbf{Center panel:} Dendrogram of hierarchical clustering with complete linkage based on one minus the correlation matrix. \textbf{Right panel:} Plot of dendrogram merge heights against the number of clusters formed at that merge. The panel, combine with the adjacent dendrogram seems to suggest that there are many clusters in the data, most of which will be singleton clusters.}}
		\label{fig:nrti_heat}
	\end{figure}

\citet{rhee2003}  study six nucleoside reverse transcriptase inhibitors (NRTIs) that are used to treat HIV-1. The target of these
drugs can become resistant through mutation, and Rhee et al.
 compare a collection of models for predicting these drugs log susceptibility, 
a measure of drug resistance based on the location of mutations. 	The design matrix $X$ comprises of $n = 1057$ rows and $p = 217$ columns. Columns were centered and scaled to have $\ell_2$-norm $n$. Columns were permuted to ensure that highly correlated columns were close to each other. Figure~\ref{fig:nrti_heat} shows some plots in an initial analysis and clustering of the data. There is a clear strong cluster, with two or three smaller, weaker ones, but many uncorrelated variables. We expect there to be many clusters in the clustering, with the vast majority of them being singleton clusters. Indeed, the gap statistic suggests that there are 161 clusters in the data. However, we set the number to $K = 150$. This ensures that the visible clusters are captured.
	
\begin{figure}[htb]
		\centering
		\includegraphics[width=110mm]{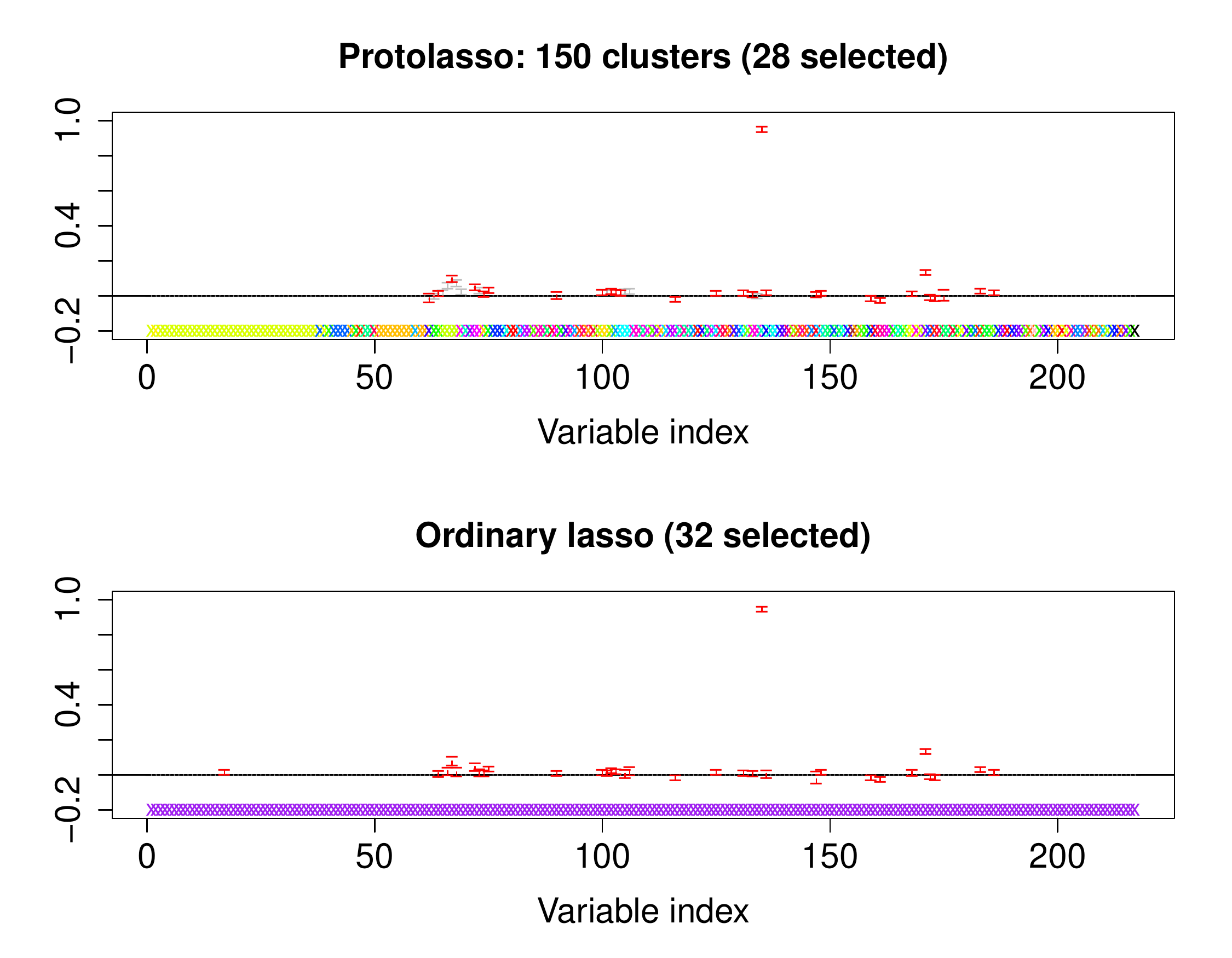}
		\caption{\emph{\textbf{Top panel:} $95\%$ post selection confidence intervals for variables in the entertained set of the \protolasso\ procedure applied to the NRTI data. Selected prototypes have red intervals; other members have grey intervals. Number of clusters in initial clustering set to $K = 150$, while regularization parameter in second stage lasso on prototypes selected via 10-fold cross validation. Clustering structure represented by coloured crosses close to horizontal axis. Singleton clusters get the same colour to reduce clutter. \textbf{Bottom panel:} $95\%$ post selection confidence intervals for variables selected by the ordinary lasso applied to the NRTI data. Regularization parameter selected using 10-fold cross validation.}}
		\label{fig:nrti_ci}
	\end{figure}

	Since $n > p$, we can estimate the variance using the least squares variance estimate. This gives $\hat{\sigma} = 0.2252$. Given this estimate, we are free to run the \protolasso\ and ordinary lasso procedures, selecting prototypes in the second stage and variables in the original matrix (respectively) via 10-fold cross validation. We then construct confidence intervals for the variables selected by the lasso and for the prototypes selected by the second stage lasso in \protolasso\ and for the other members of these selected clusters. These confidence intervals are shown in Figure~\ref{fig:nrti_ci}.
	
	We notice that the two methods select very similar entertained sets, constructing very similar confidence intervals for those variables selected. Some interesting observations:
	\begin{itemize}
		\item Both detect variable 135 and give it an interval lying quite far above everything else. This variable occurs in a singleton cluster, so \protolasso\ has nothing else to say about it.
		\item The lasso selects one variable from the large, strong group on the far left (number 17) and constructs an interval for it -- [0.0034, 0.03111]. \Protolasso\ does not consider this cluster at all.
		\item Variables 65 -- 68 occur in the same cluster. The ordinary lasso selects all but number 65 and constructs intervals for them, the interval for number 68 including zero. \Protolasso, on the other hand, gets at the same set of variables by selecting only 67 as a prototype, but then constructing intervals for all of them, because they are in the entertained set. \Protolasso\ says the $95\%$ CI for variable 68 is [0.0567, 0.0915].
		\item Something similar happens for variables 69, 72 and 73. Lasso only selects 72 and 73, while \protolasso\ selects 72 as prototype and then gets at all of them via the entertained set.
	\end{itemize}

	\section{Discussion}	
	\label{sec:discussion}
	We have introduced a coherent procedure for clustering, prototyping and subsequent analysis of datasets with groups of correlated variables. 
%	Although none of the components are particularly novel, the combination and integration of these steps have not been encountered as seamlessly before. 
	The biggest selling point of our procedure is our use of the post-selection framework of \cite{LeeSun2TaylorPostSel} to obtain \textit{exact} p-values in subsequent (i.e. post prototyping) regression and/or testing.
	Furthermore, we show how the recently proposed knockoff procedure of \cite{BC2014} can be adapted to our \protolasso\ procedure, guaranteeing FDR control with reasonable power to detect variables in true signal clusters.  
	%One of   contributions here is to provide the glue and fudge the demarcation lines between different phases, knitting the distinct steps more closely into a consistent whole.
	Subsequent research will include further development and study of the \prototest\ procedure, as marginal testing of correlated features is an important and challenging problem, for example in computational biology.
\medskip

{\bf Acknowledgments:}  We would like to thank Peter Buhlmann and Emmanuel Candes for helpful conversations. Robert Tibshirani was supported by NSF grant DMS-9971405 and NIH grant N01-HV-28183.
	\bibliographystyle{agsm}
	
	\bibliography{cluster}
	\section*{Appendix  A: \textit{A gap statistic for the automatic selection of the number of clusters in hierarchical clustering}}
	%	\label{sec:gapstat}
	
	\begin{figure}[hbtp]
		\centering
		\includegraphics[width=140mm]{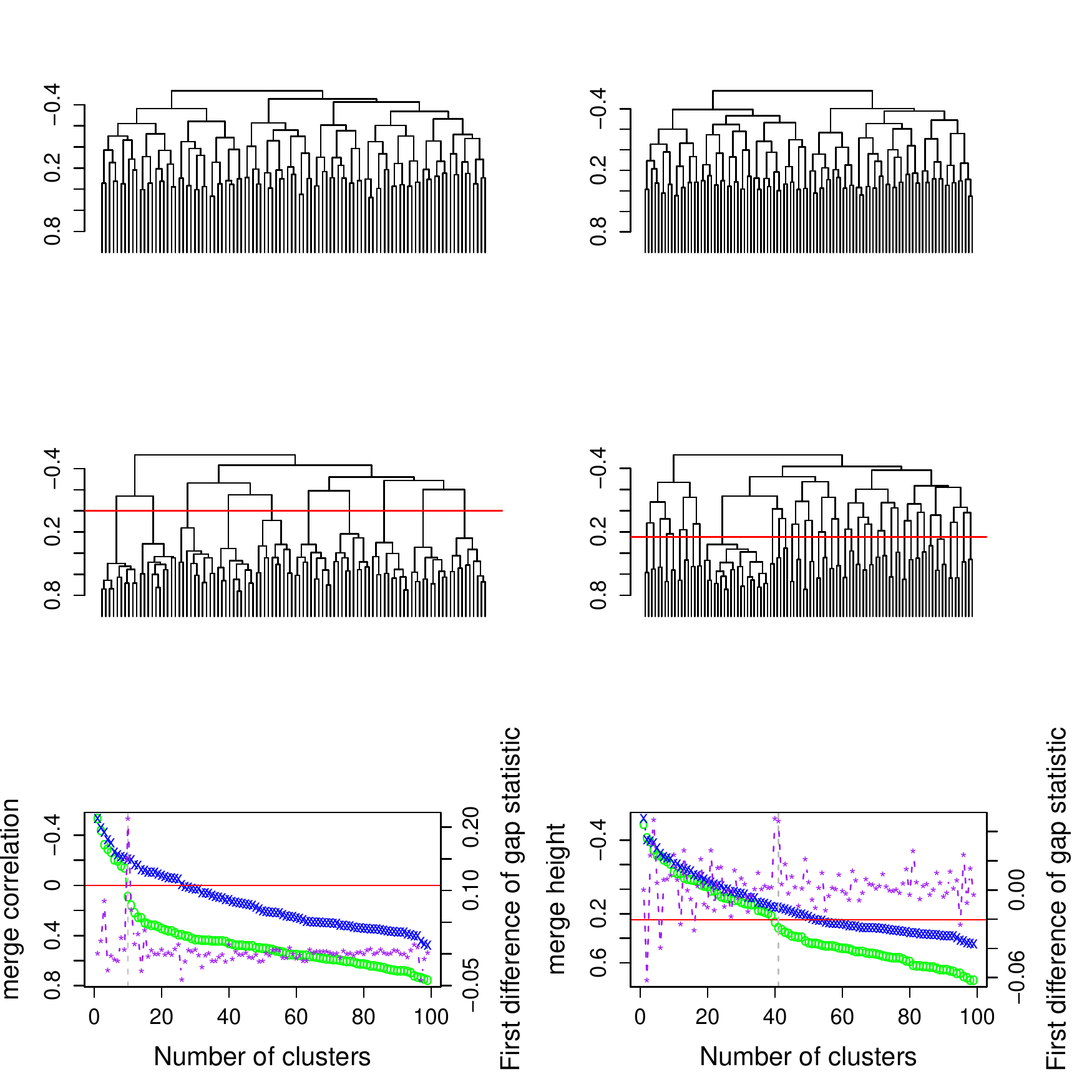}
		\caption{\emph{Dendrogram plots and plot of dendrogram cut heights against number of clusters present at that height. \textbf{Left column:} $X$ has $n = 50$ rows and $p = 100$ columns, divided into 10 clusters. Columns within clusters share pairwise correlations of $\rho = 0.5$. Columns in different clusters are uncorrelated. \textbf{Right column:} $X$ has $n = 50$ rows and $p = 100$ columns. One cluster of size 20 has pairwise correlation $\rho = 0.5$; 40 further pairs of correlated columns make up the rest. \textbf{Top panel in each column}: Hierarchical clustering dendrogram (complete linkage) for a matrix $U$, where each column has uniformly generated replicates in the range between the minimum and maximum values of the corresponding column of $X$. \textbf{Centre panel of each column:} Hierarchical clustering dendrogram (complete linkage) for the matrix $X$. \textbf{Bottom panel of each column:} Plot of dendrogram cut heights against the number of clusters present at that height. Purple curves have their scales on the right hand vertical axis. These represent the first difference of the curve obtained from the difference between the blue and green curves (the ``gap"). Note the large spikes in the purple curves around the true number of clusters.  Red horizontal lines are representative cuts that would produce the correct  clustering. Vertical dashed grey lines show the true number of clusters.}}
		\label{fig:gap_stat1}
	\end{figure}
	
	To estimate the number of clusters, we use a gap statistic procedure very similar to that of \citet{gapstat}.  We give details here.
	Figure~\ref{fig:gap_stat1} reveals the intuition behind our idea. Each row of the plot presents results for a different dataset. Details are discussed in the caption. 
	
	For each dataset with correlated variables, $X$, we present two dendrograms. The center panel shows the hierarchical clustering dendrogram for the original $X$. In each case we see fairly clearly the range of the heights at which to cut the dendrograms so as to produce an optimal clustering -- we find a place where the vertical distances between nodes and their immediate children are large. If we plot the height at which dendrogram merges occur, as a function of the number of clusters formed with that merge (green curves, bottom panels), we should she a sudden decrease as we merge to the correct number of clusters. We see this in both panels. An immediate idea is to take the first difference of this green curve (guaranteed to be negative for linkages without inversions) and find the location of the largest decrease.
	
	However, even in the absence of correlated groups, the dendrogram height curve decreases sub-linearly. This is demonstrated by generating the matrix $U$, of the same size as $X$ with elements
	\[
		U_{ij} \sim {\rm unif}(\min\{X_{1j}, X_{2j}, \dots X_{nj}\}, \max\{X_{1j}, X_{2j}, \dots X_{nj}\})
	\]
	Notice that $U$ is of the same scale as $X$, but that the correlation between columns is broken. Another way to achieve a similar aim is simply to permute the columns of $X$, each with a different randomly selected permutation. The dendrogram produced by applying the same hierarchical clustering algorithm to $U$ is shown in the top panels and the dendrogram merge heights are shown in the bottom panels in blue. Note the monotone, sub-linear decrease in the blue curves. These curves give us an indication of what the decrease in the curve looks like when there is no correlation and is a good foil to the behavior of the dendrogram curve of $X$. It is the difference in the behavior of these curves which provides information on the number of clusters. In particular, we have in mind a ``gap" statistic of the form:
	\[
		G_k = h^U_k - h^X_k
	\]
	where $k$ is the number of clusters formed at the most recent merge and $h^X_k$ and $h^U_k$ are the heights of the merge producing this number of clusters in $X$ and $U$ respectively (green and blue curves in the bottom panels of Figure~\ref{fig:gap_stat1}). Following \citet{gapstat}, we instead look at the expected value of this quantity (over the uniform distribution, fixing $X$):
	\[
		g_k = E[h^U_k] - h^X_k
	\]
	
	The expected value can be approximated using a Monte Carlo simulation with $B$ replications of $U$ being generated: $U^{(1)}, U^{(2)}, \dots U^{(B)}$ from the uniform distribution above, then computing
	\begin{equation}
		\label{eq:gapstat}
		\hat{g}_k = \frac{1}{B}\sum_{b = 1}^Bh^{U^{(b)}}_k - h^X_k
	\end{equation}
	Note that we can estimate the standard error of the gap statistic as
	\[
		\hat{{\rm s.e.}}(g_k) = \sqrt{\frac{1}{B}\sum_{b = 1}^B\left(h^{U^{(b)}}_k - \bar{h}^{U}_k\right)^2}
	\]
	where $\bar{h}^{U}_k = \frac{1}{B}\sum_{b = 1}^Bh^{U^{(b)}}_k$.
	
	From the figure it seems as though the true number of clusters reveals itself at the point where the gap between the uniform and original dendrogram heights suddenly increases. This happens because the dendrogram heights plunge suddenly for the correlated set, but does not do so for the uniform dendrogram. A good way to detect this phenomenon is to consider the first difference of the gap statistic:
	\[
		\hat{d}_k = \hat{g}_k - \hat{g}_{k-1}
	\]
	and estimating the number of clusters as
	\begin{equation}
		\label{eq:clusters}
		\hat{k} = {\rm argmax}_k \hat{d}_k
	\end{equation}
	The difference curves $\hat{d}_k$ (for a single replication $B=1$) are plotted in purple in the right panels of Figure~\ref{fig:gap_stat1}. Note the spikes in these curves at the true number of curves.
	
	%%%%% commented out
	\begin{comment}
	Although this method seems to work well for reasonably large pairwise column correlations ($\rho \geq 0.5$), often finding the true number of clusters exactly, it has been found in our simulations that it sometimes underestimates the number of clusters, especially in datasets with many smaller clusters and those where pairwise column correlations are lower. With an eye on the subsequent steps in our prototype extraction method, where we only select one prototype per cluster, we might prefer not to underestimate clusters. Fewer clusters mean fewer prototypes extracted from the first step and this increases the possibility that we might screen important signal variables at this stage. A method that seems to increase the number of estimated clusters is one that finds the maximizing index $k^* = {\rm argmax}_k \hat{g}_k$ and then estimates the number of clusters as
	\[
		{\rm max}_{k \leq k^*, \hat{g}_k > \hat{g}_{k^*} - s_{k^*}}k
	\]
	where $s_k = \hat{{\rm s.e.}}(g_k)\sqrt{1+1/B}$.
	\end{comment}
	%%%%%
	
	\subsection*{Recovery performance of the gap statistic}
	\label{sec:gapstat_perf}
	As a basic proof of concept to convince the reader of the merit of our cluster number estimation procedure, consider a small simulation study exacted on the two datasets described in the caption of Figure~\ref{fig:gap_stat1}. We considered the grid of pairwise correlation values $\rho = 0.1, 0.2, \dots 0.9$. For each of the values of $\rho$, we generated $S = 100$ replications of the pair of described datasets and applied our gap statistic procedure to each. We recorded the number of estimated clusters and a measure of cluster quality for each run. The measure of cluster quality is the Adjusted Rand Index of \citet{randindex} -- a number between 0 and 1 measuring how well the achieved clustering matches the true underlying group structure. A value of 1 occurs when the clustering matches the truth exactly; a value of 0 occurs when no pair of columns that share a group in the true grouping share any cluster in the clustering.
	
	\begin{figure}[htb]
		\centering
		\includegraphics[width=110mm]{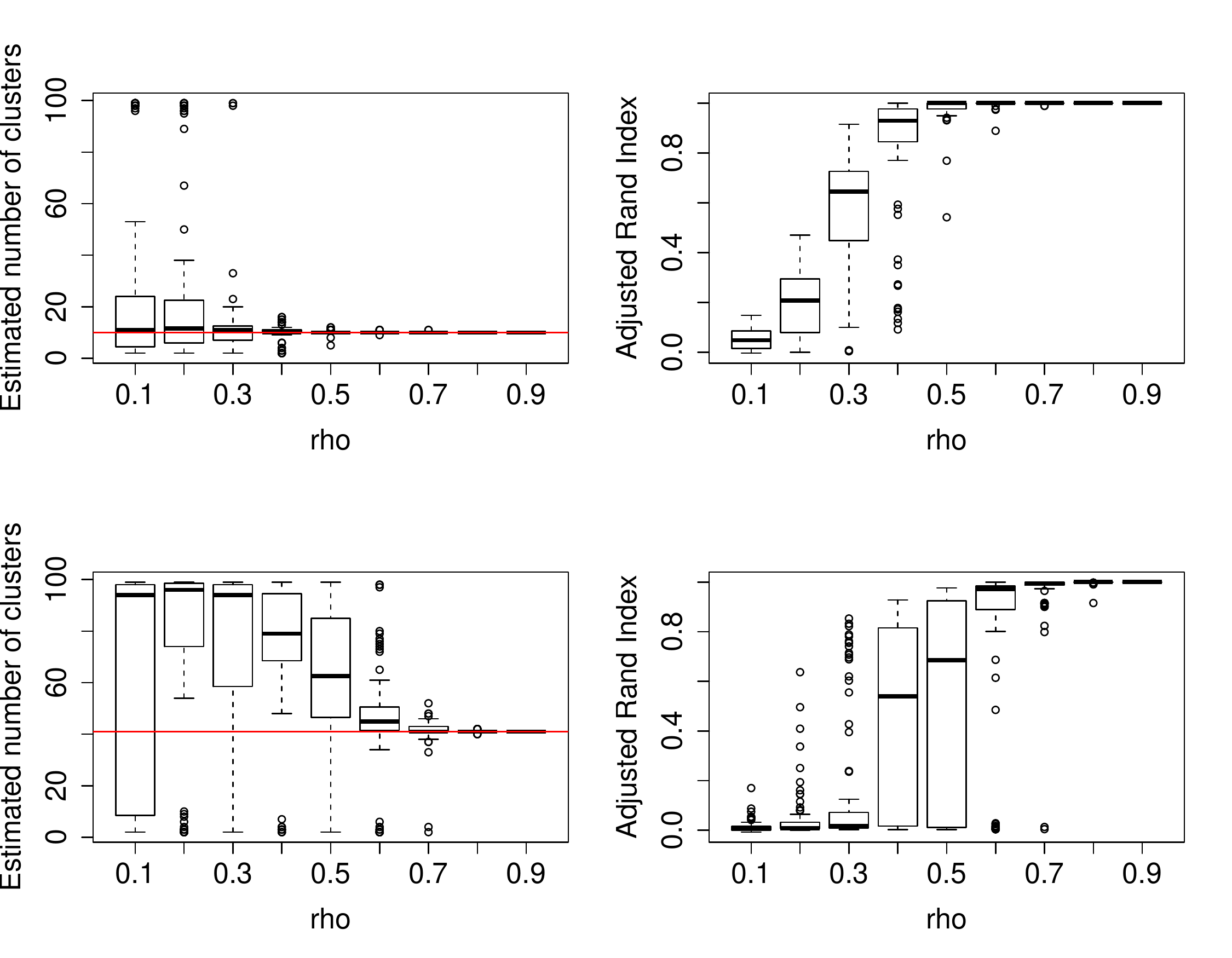}
		\caption{\emph{\textbf{Top row:} Dataset as in left column of Figure~\ref{fig:gap_stat1}. \textbf{Left panel: } Boxplots over $S = 100$ replications of the estimated cluster count using complete linkage and a gap statistic with $B = 100$ Monte Carlo simulations, as a function of the pairwise correlation amongst columns $\rho$. Horizontal red line shows the true number of clusters. \textbf{Right panel:} Boxplots over same $S = 100$ replications of the adjusted Rand Index of the clusterings so obtained, also as a function of $\rho$. \textbf{Bottom row:} Dataset the same as in the bottom row of Figure~\ref{fig:gap_stat1}. \textbf{Left panel} and \textbf{right panel:} Estimated cluster counts and adjusted Rand Indices, as in the top row, using single linkage hierarchical clustering.}}
		\label{fig:cluster_quality}
	\end{figure}
	
	We notice that the method is rather successful at recovering the correct clustering (both the numbers and cluster membership), especially at higher correlations ($\rho \geq 0.5$). At lower correlations ($\rho \leq 0.3$), there seems to be a tendency to overestimate the number of clusters (especially in the dataset in the bottom row, with its many smaller clusters -- a more difficult clustering to detect). This overestimation is not dire and is preferred to underestimation of the number of clusters. Since we pick only one prototype from each cluster, we would rather overestimate the clusters and include more prototypes in the next step than the alternative where we increase the chance of screening potentially useful signal variables at this stage.

	\section*{Appendix  B: \textit{Knockoffs with prototypes} and the exchangeability lemmas of Barber \& Cand\`{e}s}
%	\label{sec:appendix}
	\citet{BC2014} prove three lemmas serving as key ingredients in the super-martingale argument proving FDR control. In fact, the first lemma is the important property, while the latter two merely serve to aid in the proof of the first. As far as we can tell, this is the only roadblock preventing the FDR (and other results based on the super-martingale argument) in \citet{BC2014} from going through for our \textit{knockoff/prototypes} procedure. We address these three lemmas and argue that they hold for our procedure as well. Lemma 1 is stated (with mild notation modifications) as it is encountered in \citet{BC2014}.
	
	We introduce a change of notation. We let $W^{(2)}_k$ to represent the $W_j$ introduced in Section~\ref{sec:fdr}. This is merely to emphasize that these $W$-statistics stem from our knockoff/prototype procedure and depend (mostly) on $X^{(2)}$. The first lemma states that the signs of the $W^{(2)}_k$ associated with null columns are i.i.d:
	\begin{lemma} \label{lemma1}
		Let $\epsilon \in \{\pm1\}^K$ be a sign sequence independent of $W^{(2)} = (W^{(2)}_1, W^{(2)}_2, \dots, W^{(2)}_K)$, with $\epsilon_k = +1$ for all non-null prototype $\hat{P}_k$ and $\epsilon_k \stackrel{i.i.d}{\sim} \{\pm 1\}$ for null prototypes. Then
		\[
			(W^{(2)}_1, W^{(2)}_2, \dots, W^{(2)}_K) \stackrel{d}{=} (\epsilon_1\cdot W^{(2)}_1, \epsilon_2\cdot W^{(2)}_2, \dots, \epsilon_K\cdot W^{(2)}_K)
		\]
	\end{lemma}
	
	Since we assume fixed $X$, the initial clustering, and hence value of $K$, are also fixed. However, there is an extra layer of randomness in $W^{(2)}$ than in \citet{BC2014}. This randomness is injected by the prototyping step and is generated by randomness in $y^{(1)}$, which determines the set $\hat{P}$. Fortunately, $y^{(1)}$, and hence also $\hat{P}$, is independent of $y^{(2)}$, with this latter random variable generating the randomness in the knockoff procedure as understood from \citet{BC2014}. We proceed to prove the exchangeability lemmas \textit{conditional on event $\{\hat{P} = P\}$}, establishing Lemma~\ref{lemma1} above conditionally given $\{\hat{P} = P\}$. Simply integrating over the distribution of $y^{(1)}$ establishes the unconditional result as stated.
	
	The second lemma pertains to the pairwise exchangeability of features. We note that conditional on $\{\hat{P} = P\}$, if we consider the lemma for $\left[\begin{array}{c c} X^{(2)}_P & \tilde{X}^{(2)}_P \end{array}\right]$, we see that it goes through exactly as stated in \citet{BC2014}. Since we construct the knockoff $\tilde{X}^{(2)}$ from all of $X^{(2)}$ and since $X^{(2)}_P$ and $\tilde{X}^{(2)}_P$ are formed from their respective sources by selecting the \textit{same} columns, we are assured that the necessary correlation structure is retained.
	
	Lemma 3 pertains to pairwise exchangeability for the response. Again, conditional on $\{\hat{P} = P\}$, we note that $y^{(2)} \sim N(X^{(2)}\beta, \sigma^2I)$ and proof of the lemma requires
	\begin{equation}\label{lemma3proof}
		x^{(2)\top}_{P_k}X^{(2)}\beta = \tilde{x}^{(2)\top}_{P_k}X^{(2)}\beta
	\end{equation}
	where $x^{(2)}_{P_k}$ and $\tilde{x}^{(2)}_{P_k}$ are the $k^{th}$ columns of $X^{(2)}_P$ and $\tilde{X}^{(2)}_P$ respectively and $P_k$ is one of the null columns in the original $X$. Here we note that it is essential to form the knockoff matrix $\tilde{X}^{(2)}$ \textit{before} reducing to prototype matrices. We require that
	\[
		\sum_{j \not\in P} \beta_j x^{(2)\top}_{P_k}x_j = \sum_{j \not\in P} \beta_j \tilde{x}^{(2)\top}_{P_k}x_j
	\]
	
	Noting that there may be signal variables amongst the non-prototypes (i.e. $\beta_j \neq 0$ for some $j \not\in P$), we require the necessary correlation guarantees from constructing $\tilde{X}^{(2)}$ from the entire $X^{(2)}$. Our procedure as stated computes the knockoff at the correct juncture and the proof of the third lemma (conditionally on $\{\hat{P} = P\}$) follows as in \cite{BC2014}. So, conditionally on $\{\hat{P} = P\}$ we have that Lemma~\ref{lemma1} holds and hence also holds unconditionally, as argued above.
	
	Notice that this FDR guarantee holds for the prototypes individually. We are guaranteed an upper limit on the number of false nulls selected that also happen to be prototypes. If we redefine our focus to ``selecting" clusters rather than individuals prototypes, defining selection success as selecting a prototype from a cluster with at least one signal variable, we see that FDR control still holds.
\end{document}